\documentstyle[aps,tighten,preprint,graphicx]{revtex}

\begin{document}

\draft
\preprint{
\vbox{
\hbox{ADP-00-55/T435}
\hbox{PCCF RI 00-30}
}}
 
\title{Structure functions\\ for the \\ three nucleon system}

\author{F.~Bissey$^{1,2}$, A.~W.~Thomas$^1$, I.~R.~Afnan$^3$}
\address{$^1$   Special Research Centre for the
                Subatomic Structure of Matter,
                and Department of Physics and Mathematical Physics,
                University of Adelaide, 5005,
                Australia}
\address{$^2$   Laboratoire de Physique Corpusculaire,
                Universit\'e Blaise Pascal, CNRS/IN2P3,
                24 avenue des Landais, 63177 Aubi\`ere Cedex,
                France}  
\address{$^3$   School of Chemistry Physics and Earth Science,
                Flinders University,
                GPO Box 2100, Adelaide 5001,
                Australia}
  
\maketitle

\begin{abstract}

The spectral functions and light-cone momentum distributions  of protons and 
neutrons in $^3$He and $^3$H are given in terms of the three-nucleon wave 
function for realistic nucleon-nucleon interactions. To reduce computational 
complexity, separable expansions are employed for the nucleon-nucleon 
potentials. The results for the light-cone momentum distributions suggest 
that they are not very sensitive to the details of the two-body interaction,
as long as it has reasonable short-range repulsion. The unpolarised and 
polarised structure functions are examined for both $^3$He and $^3$H in order 
to test the usefulness of $^3$He as a neutron target. It is found that the 
measurement of the spin structure function of polarised $^3$H would provide a 
very clear test of the predicted change in the polarised parton distributions 
of a bound proton.
\end{abstract}

\section{Introduction}

It is well known that a polarised $^3$He target can be used as a polarised 
neutron target. The question we would like to address is how good a 
polarised neutron target it is for the determination of the neutron spin 
structure function, $g_1$, in deep inelastic scattering. There are two
questions that play a central role in resolving this problem. The first is 
the sensitivity of the light-front momentum distribution to the 
three-nucleon wave function. For this we need to calculate the spectral 
function for realistic tri-nucleon wave functions. The second question is a 
consequence of the fact that the neutron structure function is small in 
comparison with the proton structure function. This raises the question of 
the accuracy with which one can extract the polarised neutron structure 
function from $^3$He.

To examine these questions we need first to calculate the three-nucleon wave 
function for a ``realistic'' nucleon-nucleon potential. To simplify the 
problem computationally, we consider a separable expansion~\cite{HP84} of 
the Paris potential (which we call PEST)~\cite{LLR80}, that gives the same  
three-nucleon observables as the the original Paris potential in a full 
multi-channel Faddeev calculation~\cite{PKLM91,SA95}. For comparison we 
consider two other classes of potentials. The first is a rank one unitary 
pole approximation (UPA)~\cite{AR73} to the Reid Soft core 
potential~\cite{R68}. This has the property that it reproduces the position 
and residue of the poles in the $^1$S$_0$ and $^3$S$_1$-$^3$D$_1$ channels 
-- i.e., it reproduces the original potential's deuteron wave function. As a 
result, it incorporates the short range behaviour of the original 
interaction. The second is a Yamaguchi type potential with a $D$-state 
probability of 4\% and 7\%~\cite{YY54}. These potentials do not include the 
short range repulsion that is commonly present in nucleon-nucleon 
interactions.

In Sec.~2, we present the procedure used to determine the 
three-nucleon wave functions for these potentials, as well as the 
corresponding three nucleon observables. By comparing the results for these 
three classes of potential, we are able to determine the importance of short 
range correlations and the contribution of higher partial waves to the 
neutron and proton spectral functions and therefore to the light-cone 
momentum distributions. Since we will be considering both $^3$He and $^3$H, 
we have chosen to work in an isospin basis and therefore neglect the 
contribution of the Coulomb interaction to the $^3$He wave function. We do, 
however, estimate the effect of neglecting the Coulomb correction on the 
momentum distribution and therefore the structure functions.

In order to analyse the deep inelastic structure functions of $A=3$ nuclei,
we need to determine the neutron and proton spectral functions. This is 
detailed in Sec.~3. Here we compare the results for various two-body 
potentials, finding that the light-cone momentum distribution is not 
sensitive to the details of our three-nucleon wave function. In Sec.~4 we 
turn to the structure functions and examine the ratio of the structure 
function in the three-nucleon system to that in the deuteron (the EMC 
effect) for the different interactions. We also examine the possible 
implication of neglecting the Coulomb interaction in $^3$He. This opens the 
way for us to study the sensitivity of the unpolarised and polarised 
structure functions to the quark distributions in the proton and neutron and 
the possibility of extracting the neutron spin structure function from 
polarised $^3$He data. Finally, in Sec.~4 we present some concluding remarks.

\section{The three nucleon wave function}\label{3bwf}

For the three-nucleon problem we can determine the non-relativistic wave 
function  by solving the Faddeev equations exactly for any realistic 
two-body interaction. However, to simplify the computational aspects of the 
problem, with no sacrifice in the quality of the wave function, we turn to 
separable expansions that have been extensively tested~\cite{PKLM91,SA95}. 
This will result in a three-nucleon wave function that can be used to
calculate the spectral function and the light-cone momentum distribution. In 
the present section we detail the three-nucleon formalism required to 
evaluate the wave functions for $^3$He and $^3$H.

\subsection{Notation}

With the extensive literature on the Faddeev equations\cite{AT77} and their 
use in the three-nucleon system, we restrict ourselves here to a summary of 
the notation used in the present analysis. The Faddeev decomposition of the 
three-nucleon wave function is given by
\begin{equation}
\left\vert\Psi\right\rangle =\left\vert\varphi _1 \right\rangle +
\left\vert\varphi_2 \right\rangle +\left\vert\varphi_3 \right\rangle =
\{e+(123)+(132)\}\left\vert\varphi_3 \right\rangle\ .\label{Eq:1} 
\end{equation}
Here ``$e$'', ``$(123)$'' and ``$(132)$'' are members of the permutation 
group of three objects, with $e$ being the unit element ({\it i.e.} 
$e\left\vert\varphi_\alpha \right\rangle =
\left\vert\varphi_\alpha \right\rangle$) and the other two being cyclic 
permutations of $\{1,2,3\}$. The second equality results from the 
requirement that we have identical particles, the wave function is then 
invariant under any cyclic permutation of our particles. Since we have a 
system of identical fermions, the total wave function must be antisymmetric 
under the exchange of any two particles in the system. This requirement
leads to following conditions
\begin{eqnarray}
(\alpha\beta )\left\vert\varphi_\alpha \right\rangle & = & 
-\left\vert\varphi_\beta \right\rangle ,\nonumber  \\
(\alpha\beta )\left\vert\varphi_\beta \right\rangle & = & 
-\left\vert\varphi_\alpha \right\rangle ,\label{Eq:2}\\
(\alpha\beta )\left\vert\varphi_\gamma \right\rangle & = & 
-\left\vert\varphi_\gamma \right\rangle \ .\nonumber 
\end{eqnarray}
In the above equations $\alpha$, $\beta$ and $\gamma$ are indices running 
from 1 to 3, and always different from each other, and $(\alpha\beta)$ is 
again a member of the permutation group of three objects which exchange 
particles $\alpha$ and $\beta$ leaving the third one unchanged. Since we are 
dealing with a three-body problem, there will be only two independent 
momenta in the centre of mass frame. All the particles have spin and isospin 
$\frac{1}{2}$ and one must account for their orbital angular momentum. We 
briefly summarise the quantum numbers and momenta used throughout this paper:
\begin{itemize}
\item $N_\alpha$ is a set of quantum numbers describing a three body channel 
from the point of view of the particle $\alpha$, which is the spectator; the 
set is unique for each channel.
\item $\vec{\ell}_\alpha$ is the orbital angular momentum between 
particles $\beta$ and $\gamma$.
\item $\vec{L}_\alpha$ is the orbital angular momentum between particle 
$\alpha$ and the centre of mass of the system consisting of particles 
$\beta$ and $\gamma$.
\item $\vec{\jmath}_\alpha$, $\vec{\jmath}_\beta$, $\vec{\jmath}_\gamma$ 
are the spins of each particle.
\item $\vec{\imath}_\alpha$, $\vec{\imath}_\beta$, $\vec{\imath}_\gamma$ 
are the isospins of each particle.
\item $\vec{p}_\alpha$ is the momentum of particle $\alpha$ in the
centre of mass frame.
\item $\vec{q}_\alpha$ is the relative momentum of the pair of
particles $\beta$ and $\gamma$, defined as 
 $\vec{q}_\alpha =(\vec{p}_\gamma -\vec{p}_\beta )/2$.
\item $\vec{I}$ and $\vec{J}$ are respectively the total isospin and
total angular momentum of the system.
\end{itemize}

\subsection{The partial wave expansion}

We now turn to the partial wave expansion of our wave function. To
minimise the number of coupled Faddeev equations, having truncated the 
interaction to a set of partial waves, we have used the following
coupling scheme:
\begin{eqnarray*}
\vec{\jmath}_\beta+\vec{\jmath}_\gamma=\vec{s}_\alpha , & \; 
\vec{\ell}_\alpha+\vec{s}_\alpha=\vec{\bar{\jmath}}_\alpha , & \; 
\vec{\bar{\jmath}}_\alpha+\vec{\jmath}_\alpha=\vec{S}_\alpha , \; 
\vec{L}_\alpha+\vec{S}_\alpha=\vec{J} , \\
\vec{\imath}_\beta+\vec{\imath}_\gamma=\vec{\bar{\imath}}_\alpha , & 
\vec{\bar{\imath}}_\alpha+\vec{\imath}_\alpha=\vec{I}\ , & \, 
\end{eqnarray*}
which is known as the channel coupling scheme. With this coupling scheme the 
complete set of quantum number $N_\alpha$ describing a three body channel is;
$N_\alpha =\{\bar{\imath}_\alpha, s_\alpha, \bar{\jmath}_\alpha, S_\alpha, 
L_\alpha\}$. A subset of these quantum number that describe
the two-body channels is; $n_\alpha=\{\bar{\imath}_\alpha,s_\alpha, 
\bar{\jmath}_\alpha \}$, and therefore 
$N_\alpha=\{n_\alpha,S_\alpha,L_\alpha\}$. We have not included 
$\ell_\alpha$ in the set of quantum numbers since the tensor force mixes 
values of $\ell_\alpha$. This allows us to define the angular momentum and 
isospin basis as
\begin{equation}
\left| \Omega_{\ell_\alpha N_\alpha}^{JI}\right\rangle =
\left|\{L_\alpha ,
\,[(\ell_\alpha ,\,(\, \jmath_\beta ,\, \jmath_\gamma )\, s_\alpha )\, 
\bar{\jmath }_\alpha ,\, \jmath_\alpha ]\, S_\alpha \}
\, J\right\rangle\ \left | [(\, \imath_\beta ,\, \imath_\gamma )\, 
\bar{\imath }_\alpha ,\, \imath_\alpha ]\, I\right\rangle\ ,\label{Eq:3}
\end{equation}
These basis states satisfy the  following orthogonality relation 
$\left\langle \Omega_{\ell_\alpha N_\alpha }^{JI}\Big|
\Omega _{\ell_\beta N_\beta }^{JI}\right\rangle=
\delta_{\ell_\alpha,\ell_\beta}\,\delta _{_{N_\alpha ,N_\beta }}$.

We are now in a position to write the partial wave expansion of the
total three-nucleon wave function as
\begin{equation}
\left |\Psi\right\rangle =\sum _{\ell_\alpha N_\alpha}
\left|\Omega_{\ell_\alpha N_\alpha}^{JI}\right\rangle\ 
\left|{\cal U}_{\ell_\alpha N_\alpha}^{IJ}\right\rangle\ ,\label{Eq:4}
\end{equation}
where $\left|{\cal U}_{\ell_\alpha N_\alpha}^{IJ}\right\rangle$ is defined as 
the radial part of the wave function corresponding to the partial wave 
$\{\ell_\alpha,N_\alpha\}$. 

\subsection{Separable potential}

To reduce the dimensionality of the Faddeev integral equations from two to 
one, and in this way simplify the three-body wave function, we have employed 
a separable expansion of the nucleon-nucleon interaction. Our potential for 
the interaction of particles $\beta$ and $\gamma$ in a given partial wave is 
of the form\cite{AR73}
\begin{equation}
V^{n_\alpha}_{\ell_\alpha,\ell^\prime_\alpha} =
 \left| g^{n_\alpha}_{\ell_\alpha}\right\rangle 
\lambda^{n_\alpha}_{\ell_\alpha \ell^{\prime}_\alpha}\left\langle 
g^{n_\alpha}_{\ell^{\prime}_\alpha}\right\vert\ ,     \label{Eq:5}
\end{equation}
where $\left|g^{n_\alpha}_{\ell_\alpha}\right\rangle$ is a ``form factor'' and 
$\lambda^{n_\alpha}_{\ell_\alpha \ell^{\prime}_\alpha}$ is the strength of 
the potential in that partial wave. By taking 
$\ell_\alpha \neq \ell^\prime_\alpha$ we can accommodate a tensor 
interaction, as in the case of the $^3$S$_1$-$^3$D$_1$ nucleon-nucleon 
channel. The above expression for the potential is for a rank one potential. 
To incorporate higher rank potentials, we turn the strength 
$\lambda^{n_\alpha}_{\ell_\alpha \ell^{\prime}_\alpha}$ into a matrix and as 
a result $\left\vert g^{n_\alpha}_{\ell_\alpha}\right\rangle$ is a row
matrix. In resorting to separable expansions, we have taken the view that 
the expansion is a numerical procedure analogous to the use of quadratures. 
However, a low order expansion, such as the UPA or the use of a separable 
potential, is justified on the grounds that it generates the same analytic 
structure in the amplitude (i.e., bound or anti-bound state poles) as a 
corresponding realistic potential.\cite{Lo64} The use of a separable 
potential gives rise to a separable $t$-matrix that satisfies the 
Lippmann-Schwinger (LS) equation;
\begin{equation}
t_\alpha(E)=V_\alpha +V_\alpha\,G_0(E)\,t_\alpha(E)=
(1-G_0(E)\,V_\alpha )^{-1}\ V_\alpha\ ,  \label{Eq:6}
\end{equation}
with $G_0 (E)=(E-H_0)^{-1}$ the two-body Green's function. It is simple to 
show that the separable $t$-matrix in a given partial wave, resulting from 
a solution of the LS equation, is of the form
\begin{equation}
t^{n_\alpha}_{\ell_\alpha,\ell^\prime_\alpha}(E) =
\left\vert g^{n_\alpha}_{\ell_\alpha}\right\rangle
\tau^{n_\alpha}_{\ell_\alpha \ell^{\prime}_\alpha}(E)\left\langle 
g^{n_\alpha}_{\ell^\prime_\alpha}\right\vert\ ,\label{Eq:7}
\end{equation}
where the form factor $\left|g^{n_\alpha}_{\ell_\alpha}\right\rangle$ is 
identical to that used in the separable potential. The function 
$\tau^{n_\alpha}_{\ell_\alpha \ell^{\prime}_\alpha}(E)$, in a given channel, 
can be a written in matrix form as 
\begin{equation}
\left[\tau^{n_\alpha}(E)\right]^{-1} = [\lambda^{n_\alpha}]^{-1} 
- \left\langle g^{n_\alpha}\left\vert\,G_0(E)\,
\right\vert g^{n_\alpha}\right\rangle\ .        \label{Eq:8}
\end{equation}
This separability of the $t$-matrix will allow us to reduce the 
dimensionality of the Faddeev integral equations from two to one
after the partial wave expansion described in Eq.~(\ref{Eq:4}).

\subsection{The three-nucleon wave function}

Having determined the structure of the two-body amplitude, we now turn to 
the wave function for the three-nucleon system. The Schr\"odinger equation 
for this system is
\begin{equation}
(E-H_0 )\left\vert\Psi\right\rangle = V\left\vert\Psi\right\rangle =
\sum ^{3}_{\alpha =1}V_\alpha \left\vert\Psi\right\rangle
             \ .                             \label{Eq:9}
\end{equation}
This can be rewritten in a form that suggests the Faddeev
decomposition stated in Eq.~(\ref{Eq:1}), {\it i.e.},
\begin{equation}
\left\vert\Psi\right\rangle =G_0(E)\, V\left\vert\Psi\right\rangle =
\sum ^{3}_{\alpha =1}G_0(E)\, V_\alpha \left\vert\Psi\right\rangle =
\sum ^{3}_{\alpha =1}\left\vert\varphi_\alpha \right\rangle
               \ .                               \label{Eq:10}
\end{equation}
Here, $G_0(E) = (E-H_0)^{-1}$ is the three-body Green's function. We now can 
write an equation for the Faddeev components of the wave function as
\begin{equation}
\left\vert\varphi_\alpha \right\rangle =
G_0(E)\, V_\alpha \left\vert\Psi\right\rangle =
G_0(E)\, V_\alpha \left\vert\varphi_\alpha \right\rangle +
\sum _{\gamma\neq\alpha }G_0(E)\, V_\alpha 
\left\vert\varphi_\gamma \right\rangle \ .       \label{Eq:11}
\end{equation}
With the help of Eq.~(\ref{Eq:6}), the set of coupled integral equations for 
the Faddeev components of the wave function, 
$\left\vert\varphi_\alpha \right\rangle$, becomes
\begin{equation}
\left\vert\varphi_\alpha \right\rangle =G_0(E)\, T_\alpha(E) 
\left(\left\vert\varphi_\beta \right\rangle +
\left\vert\varphi_\gamma \right\rangle\right)\ .  \label{Eq:12}
\end{equation}
Here $T_\alpha(E)$ is the $t$-matrix for particles $\beta$ and $\gamma$ in 
the three-particle Hilbert space, which is related to the two-body amplitude 
considered in the last section by
\begin{equation}
T_\alpha(E) = t_\alpha(E-\epsilon_\alpha)\ ,      \label{Eq:13}
\end{equation}
where $\epsilon_\alpha$ is the energy of the spectator particle $\alpha$ in 
the three-body centre of mass.\footnote{For the three-nucleon system in a 
non-relativistic formulation, $\epsilon_\alpha = \frac{3}{4m}p_\alpha^2$, 
where $m$ is the nucleon mass.} 

In Eq.~(\ref{Eq:12}) we have a set of coupled integral equations, known as 
the Faddeev equations, for the three-body bound state. For the three-nucleon 
system, where we have identical Fermions, we take advantage of the 
anti-symmetry, as given in Eq.~(\ref{Eq:2}), and the fact that 
$(\beta\gamma)T_\alpha = T_\alpha (\beta\gamma)=-T_\alpha$, to reduce the 
Faddeev equations to
\begin{equation}
\left\vert\varphi_\alpha \right\rangle  =G_0(E)\, T_\alpha(E)\,
 \left(1-(\beta\gamma)\right)\left\vert\varphi_\beta \right\rangle 
=2\,G_0(E)\,T_\alpha(E) \left\vert\varphi_\beta \right\rangle\ ,
                                                 \label{Eq:14}
\end{equation}
with $\alpha\neq\beta$. To recast this equation into a form that will admit 
numerical solutions, we need to first partial wave decompose the Faddeev 
equations and take into consideration the separability of the two-body
amplitudes. This can all be achieved by partial wave expanding the two-body 
amplitude in three-body Hilbert space in terms of the angular momentum 
states defined in Eq.~(\ref{Eq:3})~\cite{AB77} 
\begin{eqnarray}
T_\alpha(E) &=&
\sum_{\ell_\alpha\ell^\prime_\alpha\atop N_\alpha JI}
\int\limits_0^\infty dp_\alpha\, p_\alpha^2\ 
\left|\Omega^{JI}_{\ell_\alpha N_\alpha};p_\alpha\left\rangle\ 
t^{n_\alpha}_{\ell_\alpha \ell^\prime_\alpha}(E-\epsilon_\alpha)\ 
\right\langle p_\alpha;\Omega^{JI}_{\ell^\prime_\alpha N_\alpha}\right| 
\nonumber\\ 
&=& \sum_{\ell_\alpha\ell^\prime_\alpha\atop N_\alpha JI}
\int\limits_0^\infty dp_\alpha\, p_\alpha^2\ 
\left|\Omega^{JI}_{\ell_\alpha N_\alpha};g^{n_\alpha}_{\ell_\alpha}
\left\rangle\ 
\tau^{n_\alpha}_{\ell_\alpha \ell^\prime_\alpha}(E-\epsilon_\alpha)\ 
\right\langle g^{n_\alpha}_{\ell^\prime_\alpha};
\Omega^{JI}_{\ell^\prime_\alpha N_\alpha}\right| \label{Eq:15}
\end{eqnarray}
where $\epsilon_\alpha=\frac{3}{4m}p^2_\alpha$ and
\begin{equation}
\left|\Omega^{JI}_{\ell_\alpha N_\alpha};g^{n_\alpha}_{\ell_\alpha}
\right\rangle\ = \left|\Omega^{JI}_{\ell_\alpha N_\alpha}\right\rangle\ 
\left|g^{n_\alpha}_{\ell_\alpha};p_\alpha\right\rangle\ .   \label{Eq:16}
\end{equation}
We now can write Eq.~(\ref{Eq:14}) as
\begin{eqnarray}
\left|\varphi_\alpha\right\rangle &=& 2\,G_0(E)\,
\sum_{\ell_\alpha\ell^\prime_\alpha\atop N_\alpha JI}
\int\limits_0^\infty dp_\alpha\, p_\alpha^2\ 
\left|\Omega^{JI}_{\ell_\alpha N_\alpha};g^{n_\alpha}_{\ell_\alpha}
\right\rangle\ 
\tau^{n_\alpha}_{\ell_\alpha \ell^\prime_\alpha}(E-\epsilon_\alpha)\ 
\left\langle g^{n_\alpha}_{\ell^\prime_\alpha};
\Omega^{JI}_{\ell^\prime_\alpha N_\alpha}\Big|
\varphi_\beta\right\rangle \nonumber \\
&\equiv& 2\,G_0(E)\,
\sum_{\ell_\alpha\ell^\prime_\alpha\atop N_\alpha JI}
\int\limits_0^\infty dp_\alpha\, p_\alpha^2\ 
\left|\Omega^{JI}_{\ell_\alpha N_\alpha};g^{n_\alpha}_{\ell_\alpha}
\right\rangle\ 
\tau^{n_\alpha}_{\ell_\alpha\ell^\prime_\alpha}(E-\epsilon_\alpha)\ 
X^{JI}_{N_\alpha \ell^\prime_\alpha}(p_\alpha)\ ,      \label{eq:17}
\end{eqnarray}
with the spectator function, $X^{JI}_{N_\alpha\ell_\alpha}(p_\alpha)$,  
satisfying the equation
\begin{eqnarray}
X^{JI}_{N_\alpha\ell_\alpha}(p_\alpha) &\equiv& 
\left\langle g^{n_\alpha}_{\ell_\alpha};
\Omega^{JI}_{\ell_\alpha N_\alpha}\Big|
\varphi_\beta\right\rangle \nonumber \\
&=& 2\,\sum_{\ell_\beta\ell^\prime_\beta\atop N_\beta}
\int\limits_0^\infty dp_\beta\, p_\beta^2\
Z^{JI}_{\ell_\alpha N_\alpha;\ell_\beta N_\beta}(p_\alpha,p_\beta;E)
\nonumber \\&&\qquad\qquad \qquad \times
\tau^{n_\beta}_{\ell_\beta\ell^\prime_\beta}(E-\epsilon_\beta)\ 
X^{JI}_{N_\beta\ell^\prime_\beta}(p_\beta)\ ,         \label{eq:18}
\end{eqnarray}
where
\begin{equation}
Z^{JI}_{\ell_\alpha N_\alpha;\ell_\beta N_\beta}(p_\alpha,p_\beta;E) 
\equiv \left\langle g^{n_\alpha}_{\ell_\alpha};
\Omega^{JI}_{\ell_\alpha N_\alpha}\right|\,G_0(E)\left|
\Omega^{JI}_{\ell_\beta N_\beta};g^{n_\beta}_{\ell_\beta}
\right\rangle\ ,                                  \label{eq:19}
\end{equation}
with $\alpha\neq\beta$. In Appendix~\ref{App.1} we give an explicit 
expression for $Z^{JI}_{\ell_\alpha N_\alpha;\ell_\beta N_\beta}$, for the 
coupling scheme used in the present analysis \cite{AT77,AB77}. In 
Eq.~(\ref{eq:18}) we have a set of coupled, homogeneous, integral equations 
for the spectator wave function, $X^{JI}_{N_\alpha\ell_\alpha}(p_\alpha)$, 
which we can use to construct the total wave function. Here, we note that 
the spectator wave function is only a function of the momentum of the 
spectator particle and the energy of the system, which is the binding energy 
of $^{3}$He or $^{3}$H. We now turn to the total wave function for the 
three-nucleon system. Making use of the orthogonality of the angular 
functions, $\left|\Omega^{JI}_{\ell_\alpha N_\alpha}\right\rangle$, we can 
write the total radial wave function, defined in Eq.~(\ref{Eq:4}), as
\begin{eqnarray}
    \left|{\cal U}^{JI}_{N_{\alpha}\ell_{\alpha}}\right\rangle &=&
    \left\langle \Omega^{JI}_{\ell_\alpha N_\alpha}\Big|\Psi\right\rangle
    \nonumber \\
    &=&\left\langle\Omega^{JI}_{\ell_\alpha,N_\alpha}\Big|\varphi_{\alpha} 
    \right\rangle + \left\langle\Omega^{JI}_{\ell_\alpha N_\alpha}\Big|
    \varphi_{\beta}+\varphi_{\gamma}\right\rangle \nonumber \\
    &=&\left|\eta^{JI1}_{\ell_{\alpha}N_{\alpha}}\right\rangle +
    \left|\eta^{JI2}_{\ell_{\alpha}N_{\alpha}}\right\rangle\ ,
                                                  \label{Eq:20}
\end{eqnarray}
where
\begin{eqnarray}
\eta^{JI1}_{\ell_\alpha N_\alpha}(p_\alpha,q_\alpha) &\equiv&
\left\langle p_\alpha q_\alpha\Big|\eta^{JI1}_{\ell_{\alpha}N_{\alpha}}
\right\rangle \nonumber \\
&=&\left\langle p_\alpha q_\alpha;\Omega^{JI}_{\ell_\alpha N_\alpha}
\Big|\varphi_{\alpha}\right\rangle \nonumber \\
&=&  2G_{0}(q_\alpha,p_\alpha;E)\,
       g^{n_{\alpha}}_{\ell_{\alpha}}(q_\alpha)
    \sum_{\ell_{\alpha}^{\prime}}
    \tau^{n_{\alpha}}_{\ell_{\alpha}\ell^{\prime}_{\alpha}} 
    (E-\epsilon_{\alpha})
    X^{JI}_{N_{\alpha}\ell^{\prime}_{\alpha}}(p_\alpha)\ , 
                                                      \label{eq:21}
\end{eqnarray}
with $G_0(q_\alpha,p_\alpha;E) = \left[E-\frac{1}{m}\left(q^2_\alpha +
\frac{3}{4}p^2_\alpha\right)\right]^{-1}$. The second component of
the radial wave function in Eq.~(\ref{Eq:20}) is given by
\begin{eqnarray}
\eta^{JI2}_{\ell_\alpha N_\alpha}(p_\alpha,q_\alpha) &\equiv&
\left\langle p_\alpha q_\alpha\Big|\eta^{JI2}_{\ell_{\alpha}N_{\alpha}}
\right\rangle \nonumber \\ 
&=& \left\langle p_\alpha q_\alpha;\Omega^{JI}_{\ell_\alpha N_\alpha}
    \Big|\varphi_\beta + \varphi_\gamma\right\rangle \nonumber \\    
&=& P\sum_{\ell_{\beta}N_{\beta}}\ \int\limits_{-1}^{+1} d\xi\ 
\Gamma^{JI}_{\ell_\alpha N_\alpha;\ell_\beta N_\beta} 
(p_\alpha, p^\prime_\beta;x)\, 
\eta^{JI1}_{\ell_{\beta}N_{\beta}}(p^\prime_\beta,q^\prime_\beta)
\ ,                                                  \label{Eq:22}
\end{eqnarray}
where $P= \frac{1}{2}\left[1-(-1)^{\ell_\alpha + s_\alpha +
\bar{\imath}_\alpha}\right]$, and 
\begin{equation}
p'^{\,2}_\beta = q^2_\alpha + \frac{1}{4}p^2_\alpha + q_\alpha
p_\alpha \xi\ ,\quad
q'^{\,2}_\beta=\frac{1}{4}q^2_\alpha + \frac{9}{16}p^2_\alpha -
\frac{3}{4}q_\alpha p_\alpha \xi\ ,\quad
x = -\frac{1}{p'_\beta}\left(\frac{1}{2}p_\alpha + q_\alpha \xi
\right)\ .                                          \label{Eq:23}
\end{equation}
The function $\Gamma^{JI}_{\ell_\alpha N_\alpha;\ell_\beta N_\beta}$ is 
given in Appendix~\ref{App.1}. We only observe here that the expression for 
$\Gamma^{JI}_{\ell_\alpha N_\alpha;\ell_\beta N_\beta}$ differs from that 
for $Z^{JI}_{\ell_\alpha N_\alpha;\ell_\beta N_\beta}$ by the absence of the 
separable potential form factors and the three-body Green's function. The 
normalisation of the total wave function is then given by
\begin{eqnarray}
\left\langle\Psi|\Psi\right\rangle &=&
3\langle\varphi_\alpha|\varphi_\alpha\rangle + 
6\langle\varphi_\alpha|\varphi_\beta\rangle \nonumber \\
&=& 3\sum_{\ell_\alpha N_\alpha}\left[ \left\langle\eta^{JI1}
_{\ell_\alpha N_\alpha}\Big|\eta^{JI1}_{\ell_\alpha N_\alpha}\right\rangle 
+ 2 \left\langle\eta^{JI1}_{\ell_\alpha N_\alpha}\Big|
\eta^{JI2}_{\ell_\alpha N_\alpha}\right\rangle\right]
\ .                                                  \label{Eq:24}
\end{eqnarray}
Here the sum is restricted by the two-body partial waves included in the 
Faddeev equations. Since the partial wave expansion of the total wave 
function involves an infinite sum, we need to truncate this sum such that 
the normalisation evaluated by the truncated sum, that is:
\begin{equation}
\left\langle\Psi|\Psi\right\rangle = \sum_{\ell_\alpha N_\alpha}\ 
\left\langle{\cal U}^{JI}_{N_\alpha \ell_\alpha}\Big|
{\cal U}^{JI}_{N_\alpha \ell_\alpha}\right\rangle ,  \label{Eq:25}
\end{equation}
agrees with the result of Eq.~(\ref{Eq:24}). In this way we ensure that our 
total wave function includes all the partial waves dictated by the two-body 
interaction.

\subsection{Numerical results}

As a first step in the determination of our wave function, we calculate the 
binding energy of the three-nucleon system for the class of potentials being 
considered. For the UPA to the Reid Soft core and the Yamaguchi potentials 
the interaction is restricted to the $^1$S$_0$ and $^3$S$_1$-$^3$D$_1$ 
channels. This reduces the homogeneous Faddeev equations to five coupled 
integral equations for the spectator wave function. For the PEST potentials 
the number of coupled channels depends on the rank of the interaction in a 
given channel and the number of partial waves included. To get the optimal 
representation of the Paris potential we need to have achieved convergence 
in the rank. This varies from channel to channel. In all cases the rank has
been chosen in such a way that the binding energy for a given number of 
channels has converged and is in agreement with the results of calculations 
using the Paris potential directly\cite{SA95}.
In Table~\ref{Table.1} we present the result for the binding energy for the 
three classes of potentials. For the PEST potentials we have taken the 5, 
10, and 18 channel potentials. The 18 channel calculation corresponds to 
including all nucleon-nucleon channels with $J\leq 2$. This will allow us to 
examine the contribution to the spectral function from higher partial waves. 
Here we observe that the Yamaguchi potentials overbind the three-nucleon 
system, while the UPA and PEST potentials underbind. Since the binding 
energy determines the long range part of the wave function, this difference 
allows us to examine the sensitivity of the structure functions to the 
binding energy and therefore to the tail of the wave function. A comparison 
of the PEST five channel and the UPA suggests that the difference between 
these two models is minimal. In fact, that is the case for most realistic 
potentials that do not include energy dependence. The higher partial waves 
in the PEST potential seem to have a small but significant contribution to 
the binding energy. Here again, this potential, in common with all realistic 
potentials, tends to under-bind the three nucleon system. The solution to 
this problem may involve the short-range, velocity dependence of the 
two-nucleon force \cite{Machleidt:1996km}, as well as a genuine three-body 
force \cite{BEP}.

Since we have neglected the Coulomb contribution to the energy of $^3$He, 
and our more realistic potentials under-bind the three nucleon system, we 
have chosen to adjust the strength of the $^1$S$_0$ interaction to
reproduce the experimental binding energy of both $^3$He and $^3$H. This 
procedure does not effect the deuteron wave function, but could have some 
influence on the continuum wave function in the $^1$S$_0$. In this way, we 
may estimate the error in neglecting the Coulomb energy for $^3$He, and the 
possible error in the tail of the wave function due to underbinding of the 
three nucleon system. The contribution of this correction will be discussed 
when considering the spectral functions and light-cone momentum 
distributions.

\section{Light cone momentum distribution}\label{Sec.3}

Before we proceed with the discussion of light-cone momentum distributions, 
we should establish the relation between the cross section in charged
lepton scattering and the light-cone momentum distribution. The cross 
section for the scattering of a charged lepton with a nucleus is 
proportional to the product of the leptonic tensor $L_{\mu \nu}$ with the 
hadronic tensor $W_{\mu \nu}$. For an unpolarised hadronic system of 
spin $1/2$ ({\it i.e.} free nucleon, $^3$He and $^3$H) the hadronic tensor 
has the following form \cite{GST95,LR00,AEL95}
\begin{eqnarray}
W_{\mu \nu} & = & \frac{1}{2} \sum_{S}\int d^4 x e^{iqx}
       \left\langle PS \left\vert J_\mu(x)J_\nu(0) 
       \right\vert PS \right\rangle           \label{Eq:26}\\
      & = & \left(-g_{\mu \nu}+\frac{q_\mu q_\nu}{q^2}\right)W_1  
         +\left(P_\mu-\frac{P\cdot q}{q^2}q_\mu \right)
   \left(P_\nu-\frac{P\cdot q}{q^2}q_\nu \right)\frac{W_2}{M^2}\ ,\nonumber
\end{eqnarray}
where $P$ is the four momentum of the hadronic system, $S$ is its 
polarisation and $M$ is its mass. Here, $J$ is the electromagnetic current, 
and $q$ the four momentum of the virtual photon. Finally, $W_1$ and $W_2$ 
are the form factors of the hadronic system. In deep inelastic scattering, 
one prefers to use the structure functions $F_1$ and $F_2$ instead. The 
relation between the form factors and the structure functions is the 
following
\begin{equation}
F_1 = M W_1, \quad F_2 = \frac{P \cdot q}{M} W_2\ .\label{Eq:27}
\end{equation}
The leptonic tensor for unpolarised scattering has the 
following structure \cite{GST95,LR00,AEL95}
\begin{eqnarray}
L_{\mu \nu} & = & \frac{1}{2}\sum_{s,s^{'}} 
           \bar{u}(k^{'},s^{'})\gamma_\mu u(k,s) 
           \bar{u}(k^{'},s^{'})\gamma_\nu u(k,s), \nonumber\\
           & = & 2\left(k_\mu k^{'}_\nu + k^{'}_\mu k_\nu 
           -g_{\mu \nu}k \cdot k^{'}\right)\ , \label{Eq:28} 
\end{eqnarray}
with $k$ ($k^{'}$) and $s$ ($s^{'}$) the initial 
(final) four momentum and polarisation of the lepton. \\

For polarised scattering one does not average over the initial polarisation 
and the resulting tensors then have two parts; a symmetric part, identical 
to those of Eq.~(\ref{Eq:26}) and Eq.~(\ref{Eq:28}), and a new antisymmetric 
piece that is related to the polarisation. The antisymmetric part of the 
hadronic tensor contains two new form factors, $G_1$ and $G_2$, which are in 
turn linked to two new structure functions, $g_1$ and $g_2$. 

The convolution formalism gives a prescription, valid under certain 
conditions, to link structure functions of complex hadronic systems to 
structure functions of free nucleons \cite{ScSa93,ScSa97}. In this 
formalism, the nucleon light cone momentum distribution in a nucleus plays a 
central role, in that it relates the in-medium structure function to the 
nucleon structure function. This relation takes the form of a convolution 
integral given by (see Ref.\cite{US88})
\begin{equation}
F_N^A\left(x,Q^2\right)=\int_x^\frac{M_A}{m} dy\, 
f(y)F_N\left(\frac{x}{y},Q^2\right)\ .  \label{Eq:29}
\end{equation} 
Here, $F_N$ ($F^A_N$) is the free (in medium) structure function, $f$ is the 
nucleon light cone momentum distribution, $M_A$ and $m$ are the masses of 
the nucleus and of the free nucleon respectively, finally, $x$ is the 
traditional Bjorken variable and $Q^2$ is the momentum transfer squared 
($Q^2 =-q^2$). The above relation is valid for the leading twist of the 
structure functions, which is why $f(y)$ has no $Q^2$ dependence. Another 
important assumption made in this formula is the impulse approximation, 
namely the assumption that the structure function of an off-shell nucleon is 
equal to the structure function of an on-shell nucleon. A more complete 
discussion about problems raised by this assumption can be found in 
Ref.\cite{GST95}. 

The nucleon light cone momentum distribution in a nucleus, $f(y)$, is the 
probability to find the nucleon in the nucleus with a given fraction of the 
total momentum $y(=p^+/P^+)$ of the nucleus on the light front. As a result, 
one readily see that Eq.~(\ref{Eq:29}) has a simple interpretation. The 
structure function in the medium is the sum of all possible values of the 
free nucleon structure function, weighted by the probability of finding the 
nucleon with a given momentum fraction $y$. In this section, we will show 
how to determine the light cone momentum distributions for the neutron or
proton in the three-nucleon system.  

Since the light cone momentum distribution is essentially the probability of 
finding a given nucleon with a particular fraction of the momentum 
of a nucleus, it should be related to the spectral function of the
nucleon in that nucleus. In the instantaneous frame the spectral function
is the combined probability of finding a nucleon with a given 
momentum $\vec{k}$ while the remaining nucleus is in a state $\lambda$.
We denote this spectral function by $S_\lambda (k)$. The light cone
momentum distribution is then a sum over all possible states $\lambda$,
and all possible $k$ that are compatible with the fraction of 
momentum $y$. This is given by
\begin{equation}
f(y)=\sum_\lambda \int d^4 k \left( 1+\frac{k^3}{k^0} \right) 
 \delta \left( y-\frac{k^0+k^3}{m} \right) S_\lambda (k)
                                              \ .\label{Eq:30}
\end{equation}
In some cases (see Ref.\cite{GST95}) a light cone momentum distribution
is defined for each state $\lambda$. In Eq.~(\ref{Eq:30}) the factor
$( 1+k^3/k^0)$ is called the flux factor. It is a relativistic
correction arising from the fact that we are using a light front 
formalism \cite{FS-ff,JM88}. Light cone momentum distributions, as
well as spectral 
functions, can also be defined for polarised nucleons. In the following 
section, we will concentrate on the unpolarised spectral function and 
merely state the results for the polarised nucleon spectral
function.

We note that the calculation of the nucleon momentum distributions
presented here is very similar in spirit to the pioneering work of
Ciofi degli Atti and Liuti \cite{CiofidegliAtti:1990dh}. That work
used a wave function based on variational method, rather than the
Faddeev equations. While the variational approach is designed to produce
an accurate estimate of the binding energy  of the system, one must work 
harder to obtain an equally accurate wave function. Indeed, for the 
tri-nucleon system this has led to the necessity to explicitly 
correct the proton momentum distribution, as described in Ref.\cite{CPS84}.
We are not aware of a similar correction being applied to the neutron 
momentum distribution. In any case, it appears to us that it is
worthwhile to make the calculation with a different technique. In 
addition, we can study the dependence on the assumed two-nucleon force
explicitly. 

\subsection{The spectral function}

To determine the light cone momentum distribution we need to know how 
to compute the spectral function. For the unpolarised case, the 
``diagonal spectral function'' is given by\cite{FM84,DF75}
\begin{eqnarray}
S_\lambda (k)&=&\frac{1}{2J_A +1}\sum_{\sigma_A ,\sigma} 
\left\langle\Psi,\sigma_A \left| a^{\dag}_{\sigma ,N} (\vec{k})\,
a_{\sigma ,N}(\vec{k})\right| \Psi,\sigma_A \right\rangle\ 
\delta\left(k^0-\left(m+\epsilon_\lambda-Tr_\lambda \right)\right) 
     \nonumber \\
&=&  \frac{1}{2J_A +1}\sum_{\sigma_A, \sigma\atop \sigma_b}  
\left|\left\langle \phi,\sigma_b \left|\,a_{\sigma ,N} (\vec{k})\,
\right|\Psi,\sigma_A \right\rangle\right|^2
\delta\left(k^0-\left(m+\epsilon_\lambda-Tr_\lambda 
\right)\right)                  \ .\label{Eq:31}
\end{eqnarray}
Here, $\left| \Psi,\sigma_A \right\rangle$ is the wave function of the 
initial nucleus $A$ with spin, $J_A$, and spin projection, $\sigma_A$, along 
the $z$-axis, while $\left| \phi,\sigma_b \right\rangle$ is the wave function 
of the $A-1$ system in the state $\sigma_b$. The sum over $\sigma_b$ is 
restricted to those states allowed by the energy conserving 
$\delta$-function. The operator $a^{\dag}_{\sigma ,N}(\vec{k})$ is 
the creation operator for a nucleon $N$ (proton or neutron) with spin 
projection $\sigma$ and momentum $\vec{k}$. 

In the following we will note the product 
$a^{\dag}_{\sigma ,N} (\vec{k})\,a_{\sigma ,N}(\vec{k})$ as the familiar
number density operator $\rho_{\sigma ,N}(\vec{k})$ and we will define it 
in a way similar to Ref.\cite{FGPBC90}. For example, the density of protons 
with spin $+1/2$ along the $z$--axis and momentum $\vec{\smash{p}}$, 
$\langle \rho^{+}_{p}(\vec{p})\rangle$ , in a tri-nucleon, is defined by
\begin{eqnarray}
\label{density}
\langle \rho^{+}_{p}\left(\vec{\smash{p}}\right)\rangle & = &
 \frac{1}{2}\sum_{\sigma_A} \left\langle \Psi,\sigma_A 
 \left\vert \rho^{+}_{p}\left(\vec{\smash{p}}\right) \right\vert
 \Psi,\sigma_A \right\rangle , \\
 & = &\frac{1}{2}\sum_{\sigma_A}\sum_{i = 1}^{3} 
  \int d^3\vec{\smash{q}} \left\langle \Psi,\sigma_A 
  \left(\vec{\smash{p}},\vec{\smash{q}}\right)
    \left\vert \rho^{+}_{p,i} \right\vert
  \Psi,\sigma_A \left(\vec{\smash{p}},\vec{\smash{q}}\right)\right\rangle, 
  \nonumber
\end{eqnarray}
with
\begin{equation}
\label{op-density}
\rho^{+}_{p,i}=\frac{(1+\tau_{3,i})}{2}\frac{(1+\sigma_{z,i})}{2}.
\end{equation}
In Eq.~(\ref{op-density}) one can recognise the number density, in the sense 
of Ref.\cite{FGPBC90}. The other density operators which we may use are
\begin{eqnarray*}
\rho^{-}_{p,i} & = & \frac{(1+\tau_{3,i})}{2}\frac{(1-\sigma_{z,i})}{2}, \\ 
\rho^{+}_{n,i} & = & \frac{(1-\tau_{3,i})}{2}\frac{(1+\sigma_{z,i})}{2}, \\
\rho^{-}_{n,i} & = & \frac{(1-\tau_{3,i})}{2}\frac{(1-\sigma_{z,i})}{2}. 
\end{eqnarray*}
Using the notation 
of section 2, and more specifically Eq.~(\ref{Eq:4}), we can rewrite 
Eq.~(\ref{density}) in a slightly different way, 
showing explicitly how we conduct this computation with our wave function
\begin{eqnarray}
\label{sep-density}
\langle \rho^{+}_{p}\left(\vec{\smash{p}}\right)\rangle & = &
 \frac{1}{2}\sum_{\ell_\alpha,N_\alpha,\ell_\beta,N_\beta}
 \Bigg[\left(\sum_{i,\sigma_A} \int d^2\hat{\smash{q}}
 \left\langle \Omega_{\ell_\alpha N_\alpha}^{JI},\sigma_A 
   \left(\hat{\smash{p}},\hat{\smash{q}}\right)\left\vert
\rho^{+}_{p,i} 
\right\vert \Omega_{\ell_\beta N_\beta}^{JI},\sigma_A 
 \left(\hat{\smash{p}},\hat{\smash{q}}\right) \right\rangle\right) 
\nonumber \\
 & & \qquad \times \left( \int dq\,q^2 \left\langle
  {\cal U}_{\ell_\alpha N_\alpha}^{IJ}\left(p,q\right) \Big\vert
  {\cal U}_{\ell_\beta N_\beta}^{IJ}\left(p,q\right)\right\rangle \right)
  \Bigg].
\end{eqnarray}

\subsection{The case of $^3$He}

$^3$He is one of simplest nuclei, along with $^3$H and
deuterium. It consists of 2 protons and 1 neutron. If we 
measure the light-cone momentum distribution of the neutron, the remaining
two protons can only be in a scattering state, since there is no bound 
state of two protons. On the other hand, if we measure the light cone 
momentum distribution of the proton, the remaining two nucleons are
a proton and a neutron, which can be in either a bound state, the
deuteron, or a scattering state. We will therefore study first the 
simpler case of the neutron momentum distribution and then turn to the 
more difficult proton momentum distribution. In the following equations
$\rho_N$ will mean the following 
$\sum_{i,\pm}\rho^{\pm}_{N,i}$. And whenever we omit the index $i$
it means that we implicitly sum over all three particles.

\subsubsection{Neutron in $^3$He}

In Eq.~(\ref{Eq:31}), the sum over $\sigma_b$ is constrained by the
energy conserving $\delta$-function, and for the neutron spectrum in
$^3$He this gives a scattering state for the final two protons with
the neutron off-shell. As a result the neutron does not satisfy the
on-mass-shell relation $E^2~\!=~\!\vec{\smash{p}}^2~\!+~\!m^2$. Since we are 
using a non-relativistic wave function for $^3$He we will use a 
non-relativistic approximation for the relation between the energy and the 
momentum. We then define the binding energy of the nucleus, $E$, by the 
relation $M=3m+E$, where $m$ is the mass of a nucleon. Since we are working 
with a non-relativistic wave function, we make use of the approximation
$p^0\approx m+\vec{\smash{p}}^2/(2m)$. As a result, the energy of the struck 
nucleon is $p^0_\alpha=m+E-\vec{\smash{p}}^2_\beta/(2m)-
\vec{\smash{p}}^2_\gamma/(2m)$. One then finds $p^0_\alpha=m+E-
\vec{\smash{p}}_\alpha^2/(2\mu)-\vec{\smash{q}}_\alpha^2/(2\nu)$, where 
$\nu$ is the reduced of the mass of the interacting pair and $\mu$ is their 
total mass\footnote{Note that here, in the case of two identical particles we 
have $\nu=m/2$ and $\mu=2m$.}. If we compare this result with the expression
given in Eq.~(\ref{Eq:31}), then the recoil energy $Tr$ is 
$\vec{\smash{p}}_\alpha^2/(2\mu)$, while the separation energy, $\epsilon$, is 
$E-\vec{\smash{q}}_\alpha^2/(2\nu)$. So the unpolarised spectral function for 
the neutron in $^3$He is given by
\begin{eqnarray}
\label{SFneutron}
S_n (p) & = & \frac{1}{2}\sum_{\sigma_A}\int d^3\vec{\smash{q}} 
\left\langle\Psi,\sigma_A
\left(\vec{\smash{p}},\vec{\smash{q}}\right) 
\left\vert \rho_n \right\vert 
\Psi,\sigma_A\left(\vec{\smash{p}},\vec{\smash{q}}\right) \right\rangle \\
 & &  \qquad \times \delta\left(p^0-\left(m+E-\frac{\vec{\smash{p}}^2}{2\mu}-
\frac{\vec{\smash{q}}^2}{2\nu}\right)\right) .\nonumber
\end{eqnarray}
We stress that the two forms of Eq.~(\ref{Eq:31}) are equivalent and 
should give the same results. In order to demonstrate this we computed 
the light cone momentum distribution, using Eq.~(\ref{Eq:30})
\begin{equation}
\label{fyneutron}
f_n(y)=\int d^4 k \left( 1+\frac{k^3}{k^0} \right) 
        \delta \left( y-\frac{k^0+k^3}{m} \right) S_n(k) ,
\end{equation}
with the two forms of Eq.~(\ref{Eq:31}). For the second form of this 
equation, the final state $\left\vert\phi,\sigma_b\right\rangle$ was 
taken to be a plane wave plus a pair of proton interacting in the $^1 S_0$ 
channel. This is by far the most important channel for the final 
state interaction. We found that the light cone momentum distributions 
computed with the two forms of Eq.~(\ref{Eq:31}) were identical, for all 
purpose.

For the polarised case there are two useful spectral functions
\begin{eqnarray}
\label{SFnpol+}
S_n^+ (p) & = & \frac{1}{2}\sum_{\pm}\int d^3\vec{\smash{q}} 
\left\langle\Psi^{\pm}(\vec{\smash{p}},\vec{\smash{q}}) 
\left\vert \rho_n^{\pm} \right\vert \Psi^{\pm}(\vec{\smash{p}},
\vec{\smash{q}}) \right\rangle \nonumber \\
 & & \qquad \times \delta\left(p^0-\left(m+E-\frac{\vec{\smash{p}}^2}{2\mu}-
\frac{\vec{\smash{q}}^2}{2\nu}\right)\right) ,\\
\label{SFnpol-}
S_n^- (p) & = & \frac{1}{2}\sum_{\pm}\int d^3\vec{\smash{q}} 
\left\langle\Psi^{\pm}(\vec{\smash{p}},\vec{\smash{q}}) 
\left\vert \rho_n^{\mp} \right\vert \Psi^{\pm}(\vec{\smash{p}},
\vec{\smash{q}}) \right\rangle \nonumber \\
 & & \qquad \times \delta\left(p^0-\left(m+E-\frac{\vec{\smash{p}}^2}{2\mu}-
\frac{\vec{\smash{q}}^2}{2\nu}\right)\right) .
\end{eqnarray}
These spectral functions are, respectively, for a neutron with spin parallel 
or antiparallel to the spin of the nucleus. The ``$+$'' designates a 
positive projection of the spin of either the neutron or the nucleus on 
the $z$--axis, and the ``$-$'' a negative projection. In the same way as we 
obtain $f_n(y)$ we can calculate the quantities, $f^+_n(y)$ and $f^-_n(y)$, 
just by inserting the correct spectral functions. Then one can form the 
useful quantity $\Delta f_n(y)=f^+_n(y)-f^-_n(y)$, which is the equivalent
of $f_n(y)$ for polarised structure functions. 

\subsubsection{Proton in $^3$He}

In the case of the proton we have two possibilities for the final state, 
so we also have two spectral functions. The first state is a scattering 
state similar to the final state encountered in the neutron case, with which 
it shares the formula for $p^0$. The second possible final state is made of 
a scattered proton and a deuteron. We can find the form of the 
proton energy in the same way we did for the scattering state, only it is 
now much more simple as we have only two particles in the final state and 
not three. With the same non relativistic approximation as before, one 
easily finds that in this case: 
$p^0_\alpha =M-M_d -\vec{\smash{p}}_\alpha^2/(2M_d)$, where $M_d$ is the 
deuteron mass. Defining the binding energy of the deuteron, $E_d$, in same 
way we did for the tri-nucleon we have $M_d=2m+E_d$ and finally, $p^0_\alpha =
m+E-E_d -\vec{\smash{p}}^2_\alpha/(2M_d)$. So we will have two spectral 
functions, $S^s_p(p)$ (scattering state) and $S^d_p(p)$ (deuteron state).
\begin{eqnarray}
\label{SFprotonS}
S^s_p (p) & = & \frac{1}{2}\sum_{\pm} \int d^3\vec{\smash{q}} \left\langle\Psi^{\pm}
 \left(\vec{\smash{p}},\vec{\smash{q}}\right) 
 \left\vert \rho_p \right\vert 
 \Psi^{\pm}\left(\vec{\smash{p}},\vec{\smash{q}}\right) 
 \right\rangle \nonumber \\
 & & \qquad \times \delta\left(p^0-\left(m+E-\frac{\vec{\smash{p}}^2}{2\mu}-
 \frac{\vec{\smash{q}}^2}{2\nu}\right)\right) ,\\
\label{SFprotonD}
S^d_p (p) & = & \frac{1}{2}\sum_{\pm} \int d^3\vec{\smash{q}} \left\langle\Psi^{\pm}
 \left(\vec{\smash{p}},\vec{\smash{q}}\right) 
 \left\vert \rho_p \right\vert 
 \Psi^{\pm}\left(\vec{\smash{p}},\vec{\smash{q}}\right) 
 \right\rangle \nonumber \\
 & & \qquad \times \delta\left(p^0-\left(m+E-E_d-
 \frac{\vec{\smash{p}}^2}{2M_d}\right)\right) .
\end{eqnarray}
As in Eq.~(\ref{SFnpol+}) and Eq.~(\ref{SFnpol-}) the ``$+$'' and ``$-$'' indicate the nuclear 
spin projection on the $z$--axis. 

In term of these spectral functions we can write the light cone momentum 
distribution of the proton
\begin{equation}
\label{fyproton}
f_p(y)=\frac{1}{2}\int d^4 k \left( 1+\frac{k^3}{k^0} \right) \delta 
\left( y-\frac{k^0+k^3}{m} \right)\left(S^s_p (k)+S^d_p (k)\right) .
\end{equation} 
In the preceding equation we introduced a factor one-half because there are 
two protons in a $^3$He nucleus. Without this coefficient $f_p$ would be 
normalised to $2$ instead of $1$. In the same way we did for the neutron we 
can extract polarised spectral functions, $S^{\lambda\pm}_p$, for the proton 
by using a polarised density $\rho^{\pm}_p$ in combination with the right 
polarisation of the wave function. One can then get $f^{\pm}_p$ by 
applying Eq.~(\ref{fyproton}), with the appropriate polarised spectral functions 
and in the end compute $\Delta f_p(y)=f^+_p(y)-f^-_p(y)$.

\subsection{Results} 

\begin{figure}
\centering\includegraphics[width=8.6cm,clip=true]{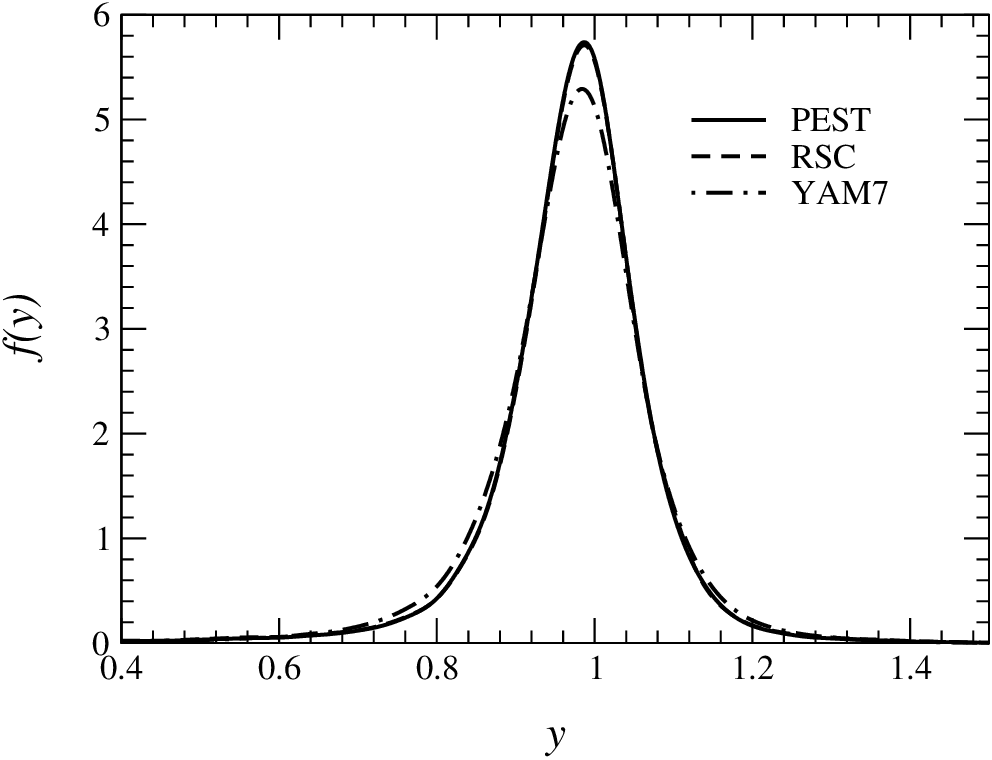}
 \caption{Neutron light cone momentum distribution in $^3$He for 
various potentials.}
 \label{Ndist}
\end{figure}

\begin{figure}
\centering\includegraphics[width=8.6cm,clip=true]{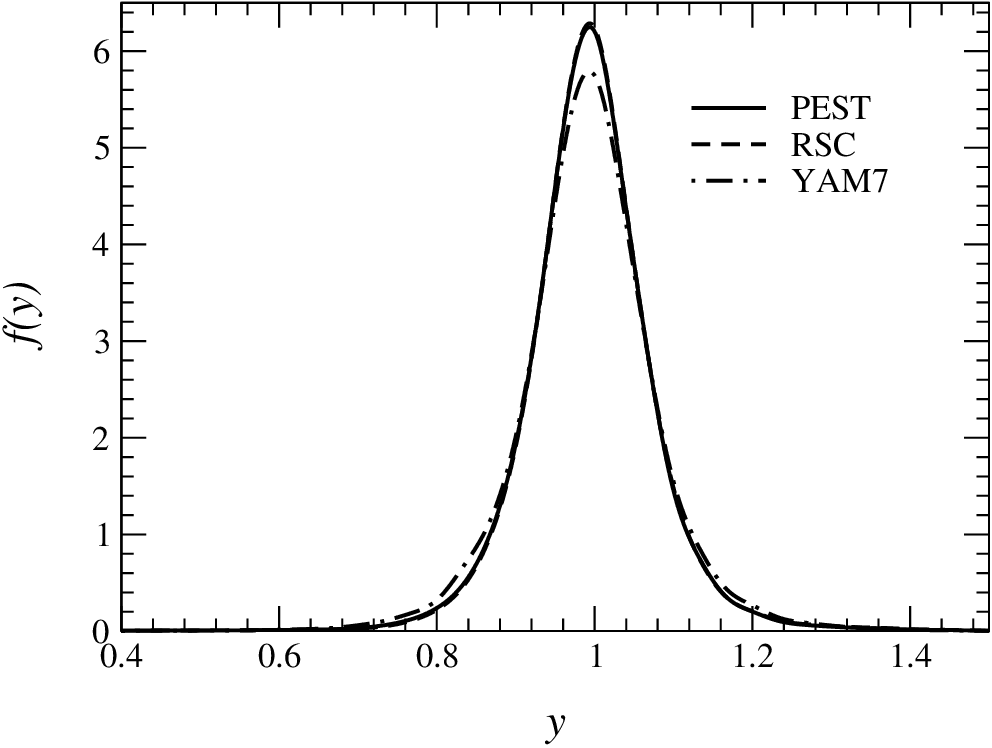}
 \caption{Proton light cone momentum distribution in $^3$He for 
various potentials.}
 \label{Pdist}
\end{figure}

Using the formalism presented above, we have computed light cone momentum 
distributions for some of our three nucleon wave functions. For all those 
distributions we used only the first $42$ three body channels. This 
because the computation of the polarised distributions involves 
some complicated matrix elements. However for all these 
wave functions the $42$ first channels add up to more than $99\%$ of the 
total, so one can safely assume that the contribution of the rest of the 
channels is negligible. For the unpolarised distribution the matrix 
elements are quite simple, so one can easily check, in this case,
that the contribution from higher channels is indeed small. We compared 
the light cone momentum distribution for a proton and a neutron in $^3$He 
for respectively $42$ and $130$ channels and found that for all purpose 
they were indistinguishable. For the PEST potential we also compared wave 
functions including $5$ and $18$ three-body channels and found that they 
were also indistinguishable. In Figs.\ref{Ndist} and \ref{Pdist} we show 
the proton and neutron light cone momentum distributions for our potentials 
(PEST, RSC and YAM7). The light cone momentum distributions given by the RSC 
and PEST potentials are almost indistinguishable and they cannot be 
separated on these figures. The YAM7 potential, however, shows some 
differences, probably because this potential does not include short range 
repulsion. It is also important to note that to have consistent results one 
needs to use a deuteron wave function computed with the same potential as 
the three nucleon system. 

\begin{figure}
\centering\includegraphics[width=8.6cm,clip=true]{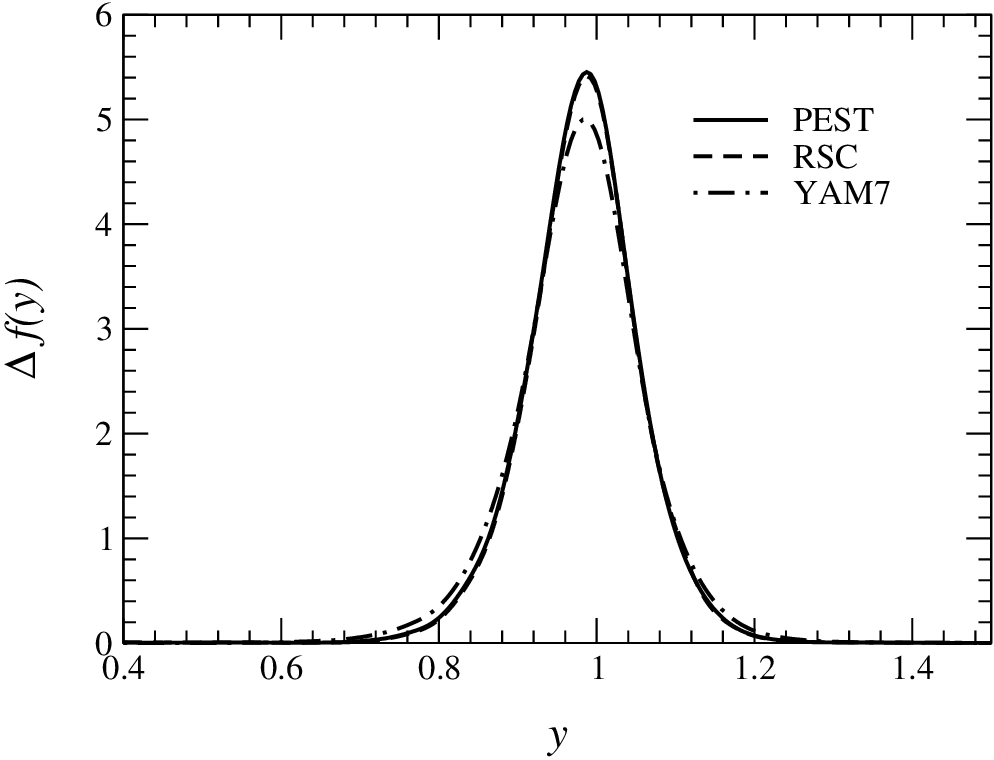}
 \caption{Neutron polarised light cone momentum distribution in $^3$He for 
various potentials.}
 \label{dN}
\end{figure}

\begin{figure}
\centering\includegraphics[width=8.6cm,clip=true]{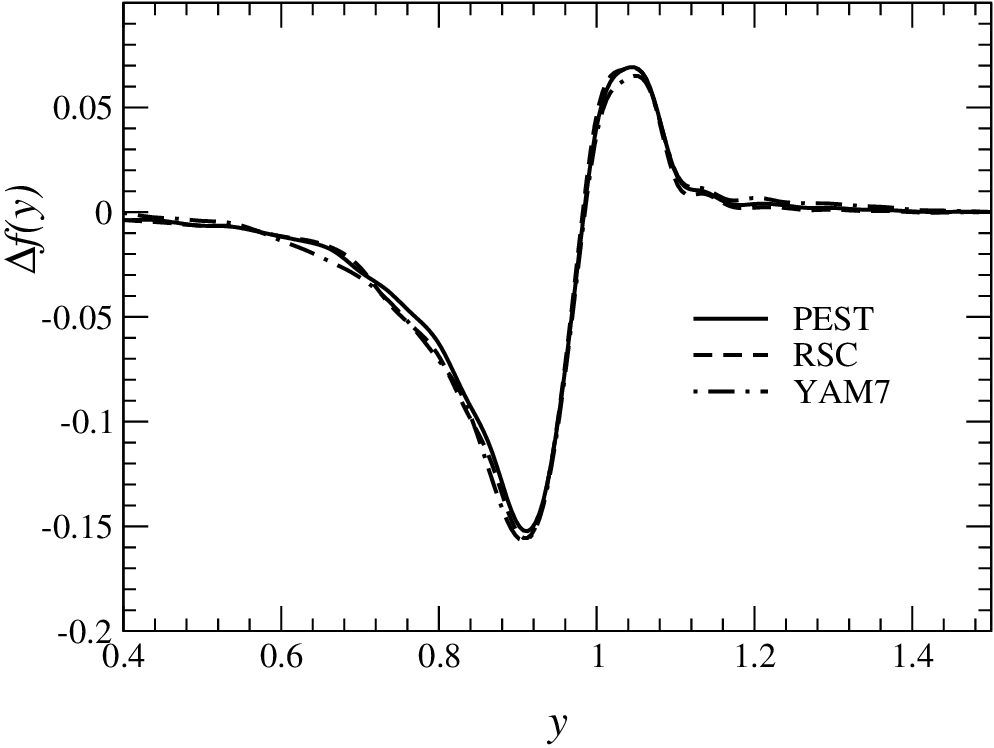}
 \caption{Proton polarised light cone momentum distribution in $^3$He for 
various potentials.}
 \label{dP}
\end{figure}

In Figs.\ref{dN} and \ref{dP} we show the proton and neutron polarised light 
cone momentum distributions for the same potentials used in 
Figs.\ref{Ndist} and \ref{Pdist}. The polarised neutron light cone
momentum distribution shows the same behavior and is similar in size
to its unpolarised counterpart. However, for the proton the polarised
momentum distribution is far smaller than its unpolarised counterpart.
In this case all the potentials gives very similar results.
We note that one can extract more information from the polarised momentum 
distributions. While in the unpolarised 
case the distributions are normalised to one, in the polarised case they are 
normalised to the polarisation of the given nucleon. From Ref.\cite{FGPBC90} 
one can compute these polarisations analytically in terms of the $S$, $S'$ 
and $D$ waves probabilities (neglecting the small contribution of the 
$P$ waves). 
One can compute those probabilities from the wave function 
and then compare them 
with the values extracted from the momentum distributions. From 
Ref.\cite{FGPBC90} we have the following relations
\begin{eqnarray}
\label{POL}
n^+ & = \int dy f_n^+(y) = & 1-\frac{1}{3}\left(P(S')+2P(D)\right) , \\
n^- & = \int dy f_n^-(y) = & \frac{1}{3}\left(P(S')+2P(D)\right) , \\
p^+ & = \int dy f_p^+(y) = & \frac{1}{2}-\frac{1}{6}\left(P(D)-P(S')\right) 
, \\
p^- & = \int dy f_p^-(y) = & \frac{1}{2}+\frac{1}{6}\left(P(D)-P(S')\right).
\end{eqnarray} 
In Table \ref{Tpola1} we compare the numerical values of these two 
expressions in $^3$He, for our various potentials. The results in quite good
agreement, with the small discrepancies arising from numerical errors in
the computation of many nested integrals. (Note, for example, that the
overall normalisation is correct to about 0.06\%.) In Table \ref{Tpola2} we 
make the same comparison but with wave functions in which we have adjusted 
the binding energies to the experimental values.

\section{Structure functions}

\subsection{Introduction}

In the incoherent impulse approximation, the structure function of a nucleus 
is the sum of the contributions from all its constituents. As we have 
already said in the previous section, the convolution formalism gives a way 
to link the in-medium structure functions to the free ones. This formalism, 
however has some limitations, especially at small Bjorken $x$, where other 
physics, like multiple scattering, becomes important. It is also only valid 
in the Bjorken limit, as the convolution formalism itself does not depend on 
$Q^2$. In unpolarised scattering this formalism is a good tool to i
nvestigate the EMC effect \cite{Au83}, so we will use our previous results 
to study this effect in the three nucleon system. Another interesting result 
from the previous section is the fact that in $^3$He, the proton 
polarisation ({\it i.e.} $\Delta_p = p^+-p^-\approx -2\%$) is very small and 
negative, while the neutron polarisation ({\it i.e.} $\Delta_n = 
n^+-n^-\approx 87\%$) is quite big. This is also clear 
from Figs.\ref{dN} and \ref{dP}. This means that the neutron carries most of 
the spin of $^3$He, so, at least for polarised scattering, this nucleus 
should be a good approximation to a pure neutron target. The same argument 
is valid for the proton in $^3$H. Since we already have a free proton target 
this may appear less interesting at first sight. On the other hand, it 
provides an ideal way to study the effect of the nuclear medium on the spin 
structure of a bound nucleon.  

\subsection{Unpolarised structure function and EMC effect}

As we explained at the beginning of the previous section, in unpolarised 
deep inelastic scattering of a charged lepton on a nuclear target, all 
the target information is included in the two structure functions $F_1$ 
and $F_2$. In a simple quark model those functions have the following 
form \cite{GST95,AEL95}
\begin{eqnarray}
\label{eqF1}
F_1(x,Q^2) & = & \frac{1}{2}\sum_q e^2_q q(x,Q^2), \\
\label{eqF2}
F_2(x,Q^2) & = & 2xF_1(x,Q^2)=x\sum_q e^2_q q(x,Q^2).
\end{eqnarray}
In these expressions $q(x)$ is the distribution of quarks of flavour $q$ 
and electric charge $e_{q}$. The relation between $F_{1}$ and $F_{2}$ 
implies that the partons have spin 1/2 and no transverse momentum in the 
infinite momentum frame. A more general relation between $F_{1}$ and 
$F_{2}$ \cite{GST95} is  
\begin{equation}
\label{eqF1toF2}
F_{2}(x)=2xF_{1}(x)\frac{1+R}{1+2xm_{N}/\nu } , 
\end{equation}
where $R$ is the ratio of the cross section for absorbing a 
longitudinal photon to that for a transverse photon. 

Given the relation 
between $F_1$ and $F_2$, most studies concentrate on the 
latter. The convolution formula between the free and in medium $F_2$ 
structure functions \cite{GST95,US88} is
\begin{equation}
\label{convolF2}
\tilde{F}^N_2(x,Q^2)=\int_x^\frac{M_A}{m} dy f_N(y)F^N_2
\left(\frac{x}{y},Q^2\right).
\end{equation}
Hence the $F_2$ structure function of a nucleus of mass number $A$ 
and proton number $Z$ is given by
\begin{equation}
\label{F2A}
F_2^A(x,Q^2)=\int_x^\frac{M_A}{m} dy \left( Z f_p(y)F_2^p
\left(\frac{x}{y},Q^2\right) +
(A-Z) f_n(y)F_2^n\left(\frac{x}{y},Q^2\right) \right).
\end{equation} 

In comparing the $F_2$ structure functions on various targets, 
the European Muon Collaboration (Aubert et al. \cite{Au83}) 
discovered what is now called the ``EMC'' effect. We define a theoretical 
EMC ratio as the ratio of the $F_2$ structure function of the nucleus to 
the sum of the free structure functions of the nucleons in this nucleus:
\begin{equation}
\label{R_t}
R_t=F_2^A/(ZF_2^p+(A-Z)F_2^n)
\end{equation}
On the other hand, it is more common to compare  
the ratio of the $F_2$ structure function of the nucleus to that of 
deuterium: 
\begin{equation}
\label{R_x}
R_x=(F_2^A/A)/(F_2^D/2)
\end{equation}
This should be close to $R_t$ if 
the deuteron is a quasi-free system of a proton and a neutron and if the 
nucleus studied is symmetric, or almost, in its content of neutrons and 
protons. $^3$He and $^3$H are highly asymmetric nuclei, as their content in 
one type of nucleon is twice as much as the other. To take this into 
account, it is common to an isosymmetric correction
so that the ratio studied is \cite{US88}:
\begin{equation}
\label{Rexp}
R_A(x,Q^2) = \frac{F_2^A(x,Q^2)}{F_2^D(x,Q^2)}I(x,Q^2)
\end{equation}
with
\begin{equation}
\label{Iso}
I(x,Q^2) = \frac{F_2^p(x,Q^2)+F_2^n(x,Q^2)}{ZF^p_2(x,Q^2)+(A-Z)F_2^n(x,Q^2)}.
\end{equation}

\begin{figure}[t]
\centering\includegraphics[width=8.6cm,clip=true]{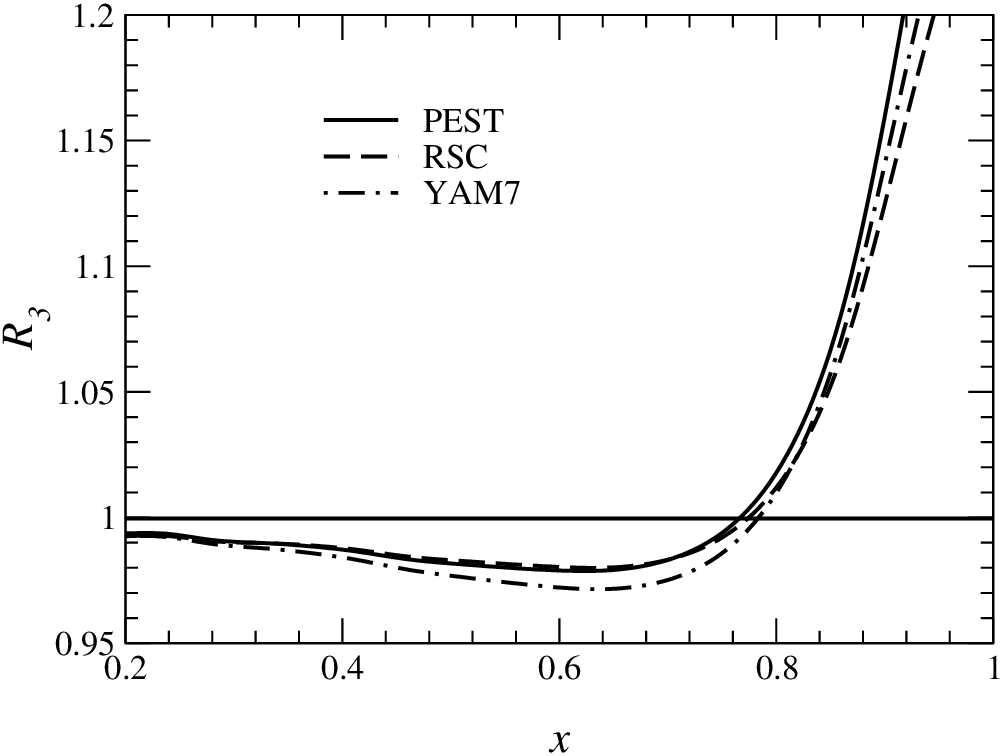}
\caption{The ratio $R_3$, given in Eq.(\ref{Rexp}), for $^3$He, 
at $Q^2=10$GeV$^2$, calculated for 
various potentials using the CTEQ5 quark distributions.}
\label{phe-iso}
\end{figure}

\begin{figure}
\centering\includegraphics[width=8.6cm,clip=true]{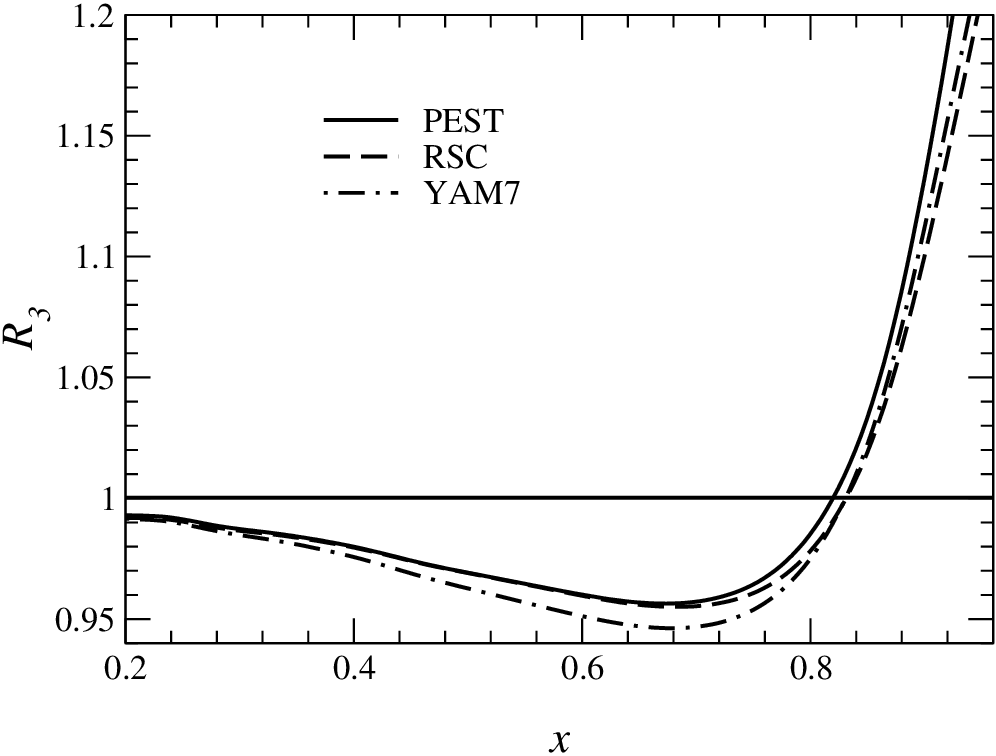}
\caption{The ratio $R_3$, given in Eq.(\ref{Rexp}), for $^3$H, 
at $Q^2=10$GeV$^2$, calculated for 
various potentials using the CTEQ5 quark distributions.}
\label{ptr-iso}
\end{figure}

\begin{figure}
\centering\includegraphics[width=8.6cm,clip=true]{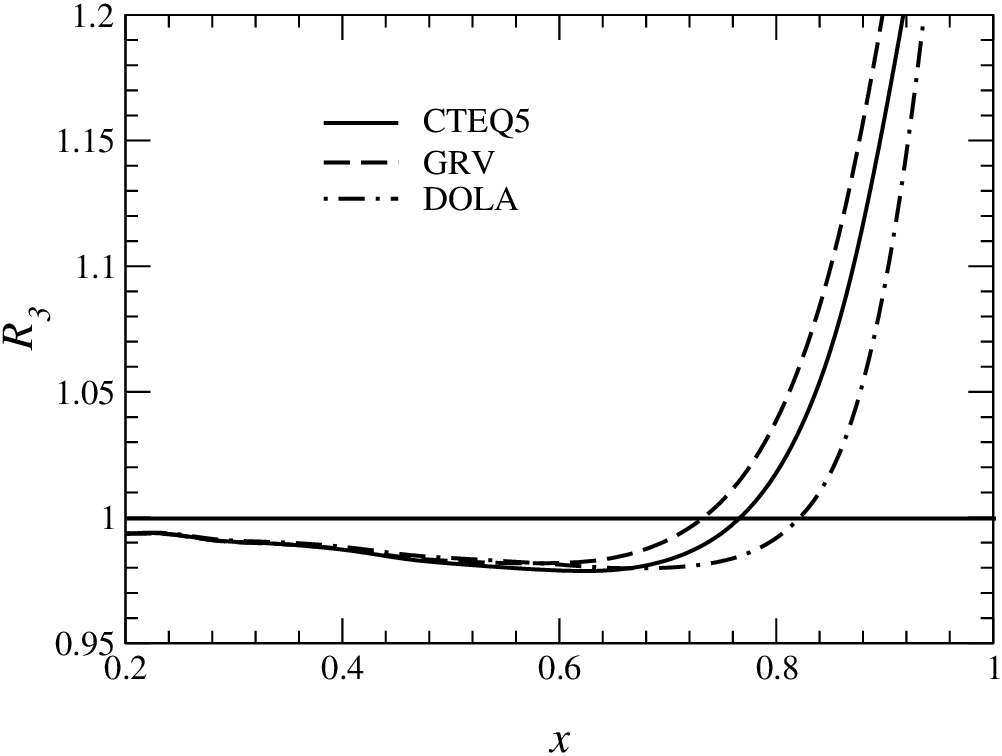}
\caption{The ratio $R_3$, given in Eq.(\ref{Rexp}), for $^3$He, 
at $Q^2=10$GeV$^2$, calculated for the 
PEST potential, using various quark distributions for the nucleons.}
\label{he-isovq}
\end{figure}

This ratio is, strictly speaking, the ratio of the EMC ratios of the nucleus 
$A$ and the deuteron. Following the same kind of procedure used in the 
previous section, one can compute the light cone momentum distribution of a 
nucleon in the deuteron. To be consistent, this ratio has to be computed 
with the same interaction for both the three nucleon system and the 
deuteron. To compute $R_A$ we used several parametrisations for the quark 
distributions:
\begin{itemize}
\item The parametrisation ``CTEQ5'' from the CTEQ collaboration 
\cite{CTEQ5}. The collaboration gives several parametrisations, but we 
mainly used the one called ``leading order'', and it will be the one used 
when we talk about the CTEQ5 parametrisation, unless explicitly stated 
otherwise.
\item The ``GRV'' parametrisation from Gl\"uck, Reya and Vogt \cite{GRV95}.
\item The ``DOLA'' parametrisation from Donnachie and Landshoff \cite{dola}.
\end{itemize}
These distributions are usually given for quarks in a proton 
and in order to 
compute neutron structure functions we used charge symmetry\footnote{With 
the exception of the DOLA distribution which gives proton and deuteron 
distributions. In this case we took the neutron as the difference between 
the deuteron and the proton.}\cite{CSrev}. 
In Figs. \ref{phe-iso} and \ref{ptr-iso}  
one can see the ratio $R_3$ for $^3$He and $^3$H, with the CTEQ5 
parametrisation at 
$Q^2=10$GeV$^2$, for the three potentials studied. In Fig.\ref{he-isovq} 
we show $R_3$ in $^3$He for the PEST potential alone but for all three quark 
distributions (again at $Q^2=10$GeV$^2$). 
We also studied the effect of adjusting 
the binding energy as described at the end of the first section but did not 
include it in Figs.\ref{phe-iso} and \ref{ptr-iso} because it would have 
confused the plot. This adjustment of the binding energy 
caused a slightly deeper 
EMC effect in both $^3$He and $^3$H and also a slightly steeper increase at 
high $x$.

\subsection{Polarised structure functions}

If one does experiments with both a polarised lepton beam and a polarised 
spin $1/2$ nuclear target, one needs two more structure functions, $g_1$ 
and $g_2$. One can perform various measurements of cross sections with 
several polarisations in order to extract those two structure functions. 
They are smaller than $F_1$ and $F_2$ and $g_2$, in particular, is often 
neglected. As we indicated in the introduction, the figures for the 
effective polarisation of the nucleons in the three nucleon system seem to 
indicate that the contribution to the nuclear spin structure functions from 
the doubly represented nucleons is severely reduced. Thus, this system 
should be a good approximation to a pure single nucleon target. At leading 
order, $g_1$ has the following form\cite{LR00,Wo89,AAC00}
\begin{eqnarray}
\label{eqg1}
g_1(x,Q^2) & = & \frac{1}{2}\sum_q e^2_q \Delta q(x,Q^2).
\end{eqnarray}
In Eq.~(\ref{eqg1}), $\Delta q$ are the polarised quark distributions. They 
involve the difference between the distributions of quarks with the same and 
opposite helicity from that of the nucleon. It is much harder to find a 
simple parton interpretation for $g_2$ \cite{LR00}. 

\begin{figure}
\centering\includegraphics[width=8.6cm,clip=true]{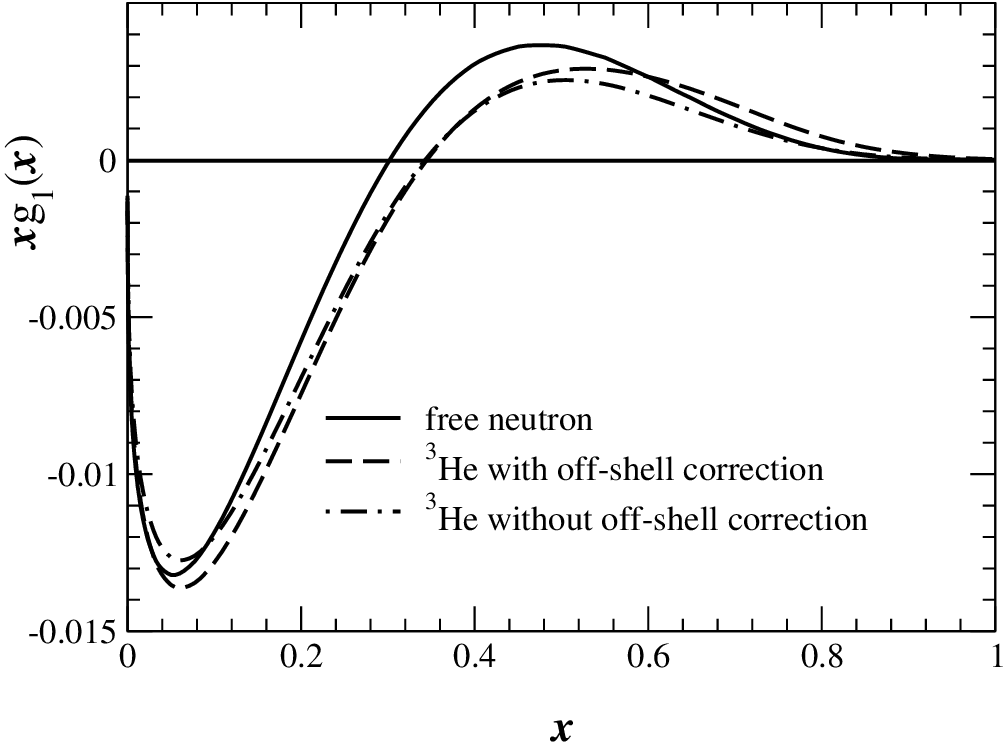}
\caption{Comparison of several calculations of $xg_1(x)$ for $^3$He, 
at $Q^2=10GeV^2$.}
\label{xg1}
\end{figure}

\begin{figure}[t]
\centering\includegraphics[width=8.6cm,clip=true]{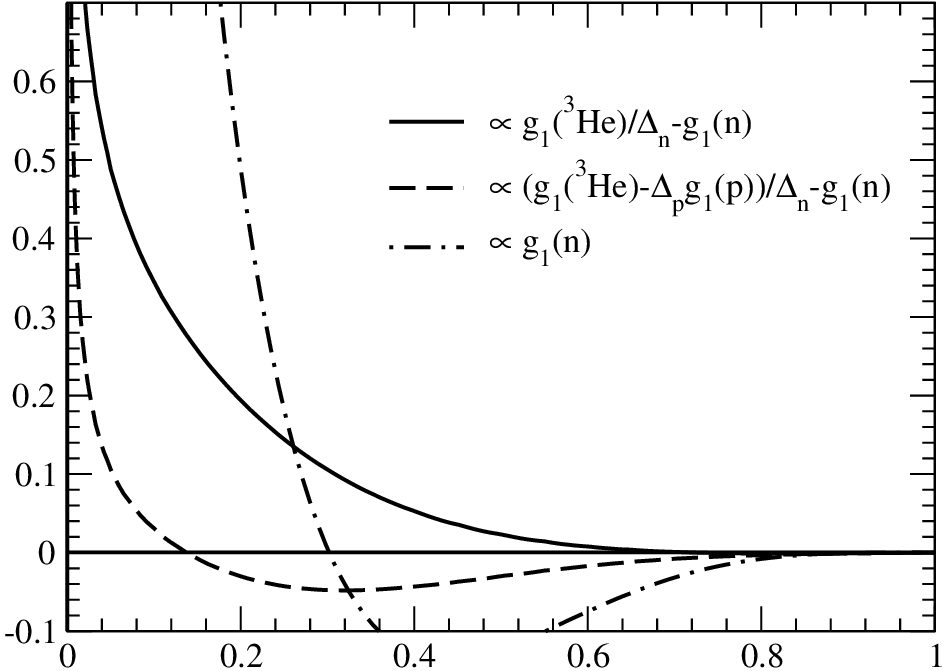}
\caption{$\Delta_g$, $\Delta^{'}_{g}$ and $g_1(n)$ at $Q^2=10GeV^2$. Note
that all three curves have been divided by $\int dx g_1 (^3He)$.}
\label{diff}
\end{figure}

\begin{figure}
\centering\includegraphics[width=8.6cm,clip=true]{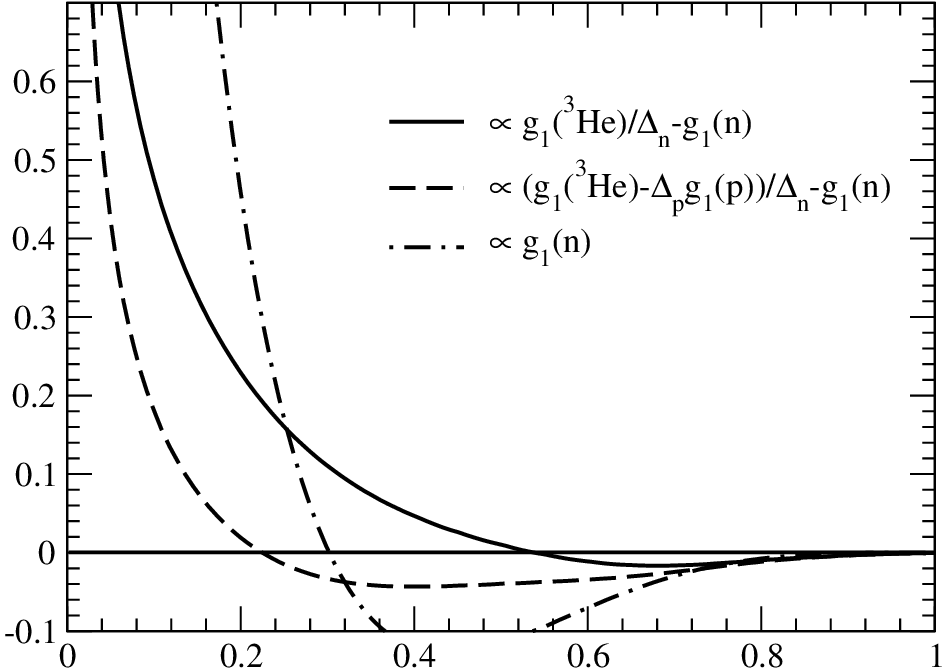}
\caption{$\Delta_g$, $\Delta^{'}_{g}$ and $g_1(n)$, including off-shell 
corrections, at $Q^2=10GeV^2$. Note that all three curves have been 
divided by $\int dx g_1 (^3He)$.}
\label{diff-os}
\end{figure}

The convolution formula relating the free spin structure function to that 
in-medium is the following:
\begin{equation}
\label{convolg1}
\tilde{g}^N_1(x,Q^2)=\int_x^\frac{M_A}{m} \frac{dy}{y} \Delta f_N(y) 
g^N_1\left(\frac{x}{y},Q^2\right). 
\end{equation}
We computed the $g_1$ structure function of $^3$He using the same three 
potentials as for $F_2$. The results from those potentials are sufficiently 
close that we will only use the results from the PEST potential hereafter.
To compute $g_1$ we mainly used the NLO ``standard scenario'' of 
Ref.\cite{GRSV96}. We also studied the impact of the off-shell correction 
from Ref.\cite{STTS99} on $g_1$. (The off-shell correction was calculated
using a local density approximation and the quark meson coupling 
model \cite{QMC} to estimate the change of the parton distributions in a 
bound nucleon.) In Fig.\ref{xg1} we show the following three curves at 
$Q^2=10GeV^2$: $xg_1(x)$ for the free neutron, as well as $xg_1(x)$ for 
$^3$He with and without the off-shell correction. As one can see, the three 
of them are close. The main complication in the extraction of $g_1$ for the 
free neutron from $^3$He is that the free proton spin structure function is 
very big compared with that of the neutron. So, while its contribution in 
$^3$He is severely reduced by the low effective polarisation, it is still not
negligible. One way to estimate the size of the contribution of the proton 
is to compare $g_1(^3He)$ with a formula often used in the experimental 
analysis (see Ref.~\cite{AEL95} for a derivation):
\begin{equation}
\label{g1approx}
g_1(^3He) \approx \Delta_n g_1(n) + 2 \Delta_p g_1(p).
\end{equation}
If the contribution of the proton to $g_1(^3He)$ is negligible, 
Eq.~(\ref{g1approx}) is equivalent to: $g_1(^3He) \approx \Delta_n g_1(n)$. 
To estimate the effect of the proton contribution in the extraction of 
$g_1(n)$, we plotted the following differences: 
\begin{equation}
\label{delta_g}
\Delta_g = \frac{g_1 (^3He)-2\Delta_p g_1 (p)}{\Delta_n} -g_1 (n) 
\end{equation}
and 
\begin{equation}
\label{delta'_g}
\Delta^{'}_{g} = \frac{g_1 (^3He)}{\Delta_n} -g_1 (n).
\end{equation}
In Figs.\ref{diff} and \ref{diff-os} we plot both $\Delta_g $ and 
$\Delta^{'}_{g}$. The second plot includes the off-shell effect of 
Ref.~\cite{STTS99}. Note that the curves have been divided by  
$\int dx g_1 (^3He)$~($\approx~-1/16$) so that one can judge the effect
on the spin sum rule. Since one ultimately wants to extract $g_1 (n)$, we 
have also plotted that with the same normalisation, so as to have an idea 
of the size of the error in the differences\footnote{We do not plot the 
ratio of structure functions because in both the neutron and $^3$He cases 
$g_1$ can be zero, leading to singularities in the plots.}. It is clear from 
both plots that one gets more accurate results by including the proton 
contribution for mid-range $x$ ($0.2 \leq x \leq 0.6$), the biggest error in 
this region occurring when the structure function crosses the $x$--axis. 
At higher $x$ ($x \geq 0.6$) the effect of Fermi motion is significant
and this will be even more important for $^3$H, below. Nevertheless, the
absolute value of the structure function is small and the corrections
have little effect on the spin sum rule. For 
smaller $x$ ($x \leq 0.2$) one clearly needs some other 
tools to accurately extract the free neutron structure function, even if the 
relative difference seems to be small. We find similar curves for other
parton distributions, such as those from Ref.\cite{GS95}. 

\begin{figure}[t]
\centering\includegraphics[width=8.6cm,clip=true]{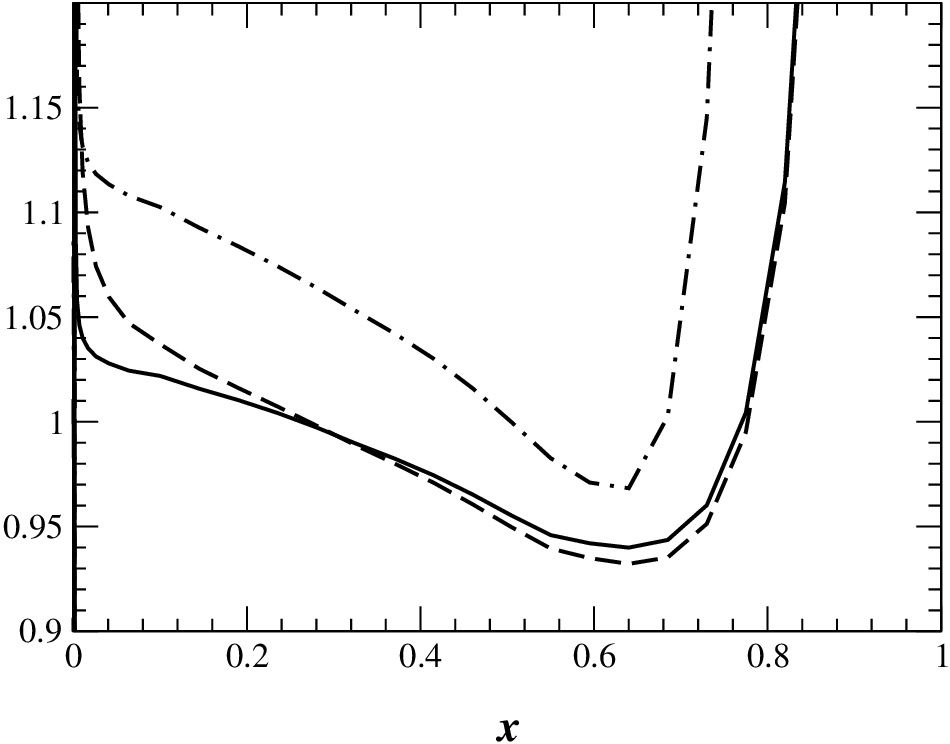}
\caption{$R_g$ and $R^{'}_{g}$ at $Q^2=10GeV^2$.}
\label{ratio-g1}
\end{figure}

In the case of tritium one can plot a ratio, as $g_1 (p)$ does not 
change sign. Therefore, to illustrate the effect of the neutron 
contribution in this case we plot: 
\begin{equation}
\label{R_g}
R_g = \frac{g_1 (^3H)-2\Delta_n g_1 (n)}{\Delta_p g_1 (p)}
\end{equation}
and 
\begin{equation}
\label{R'_g}
R^{'}_{g} = \frac{g_1 (^3H)}{\Delta_p g_1 (p)}
\end{equation}
In Fig.~\ref{ratio-g1} we show both ratios ($R_g$ is the solid line and 
$R^{'}_{g}$ is the dashed line) without including the off-shell corrections 
\cite{STTS99} as well as
$R_g$ with the off-shell corrections (dot-dashed line). In 
this figure we can clearly see that on most of the interval the contribution 
of the neutron is negligible, some difference appearing for small $x$. This
is expected simply because $g_1 (n)$ is significantly smaller than $g_1 (p)$
for most values of $x$. On the other hand, we can also see that medium 
effects seem to be quite important and that the off-shell correction makes 
an important difference. One can also see clearly the effect of Fermi motion 
at high $x$, while it would be invisible if one were to plot differences.  
It is clear from these results that from a measurement of $g_1 (^3H)$ one 
can expect to extract the size of the change in the spin structure function 
of the bound proton and one might even hope to separate the origin of 
this effect. 

\section{Conclusions}

We have computed the three-nucleon structure functions from various
two body potentials. This involved calculating wave functions, light cone 
momentum distributions and finally the structure functions. We have 
presented our prediction for the EMC effect in both $^3$He and $^3$H and 
shown that they were quite close for various two-body potentials 
and quark distributions. In addition, we saw  
that isospin breaking would have only a small effect on these findings. This
result has been used elsewhere \cite{aeq3} in a proposal to measure the
$d/u$ ratio at large $x$ at Jefferson Laboratory \cite{JLab,Pace:2000ky}. 

{}From our study of the spin structure function of $^3$He, we showed
that it possible to extract the structure function of a polarised
neutron with reasonable accuracy. However, it is necessary to account
for the contribution from the pair of protons which are not totally
unpolarised. Turning to the polarised structure function of $^3$H, 
we saw that while the experiment is extremely challenging it could also
be very valuable. In particular, one can measure the size of the medium
corrections and check experimentally the predicted
modification of the spin dependent parton distributions of the  
bound nucleon.

\section*{Acknowledgments}

This work was supported by the Australian Research Council. We would
like to thank W.~Melnitchouk for many helpful discussions.

\appendix
\section{The kernel of the homogeneous Faddeev equation}\label{App.1}

For completeness, we present in this Appendix the explicit expression for 
the kernel of the homogeneous Faddeev equation when the interaction is 
represented by a separable potential. The details of the derivation are in 
Ref.~\cite{AT77}. We have
\begin{eqnarray}
Z^{JI}_{\ell_\alpha N_\alpha;\ell_\beta N_\beta} &\equiv& 
\left\langle g^{n_\alpha}_{\ell^\prime_\alpha};
\Omega^{JI}_{\ell^\prime_\alpha N_\alpha}\right|\,G_0(E)\left|
\Omega^{JI}_{\ell_\beta N_\beta};g^{n_\beta}_{\ell_\beta}
\right\rangle \nonumber \\
&=& \frac{1}{2}\int\limits_{-1}^{+1}\,dx \frac{
\,g^{n_\alpha}_{\ell_\alpha}(q_\alpha)\,g^{n_\beta}_{\ell_\beta}
(q_\beta)}{E-\frac{1}{m}\left(p^2_\alpha+p^2_\beta +
p_\alpha p_\beta x\right)}\ \Gamma^{JI}_{\ell_\alpha N_\alpha;
\ell_\beta N_\beta}(p_\alpha,p_\beta;x)\ ,\qquad\label{eq:A1} 
\end{eqnarray}
where
\begin{eqnarray}
\Gamma^{JI}_{\ell_\alpha N_\alpha;\ell_\beta N_\beta}
(p_\alpha,p_\beta;x) &=& 
\left(\frac{p_\beta}{q_\alpha}\right)^{\ell_\alpha} 
\left(\frac{p_\alpha}{q_\beta}\right)^{\ell_\beta}
B_{N_\alpha N_\beta}\sum_{\cal L} P_{\cal L}(x)\nonumber \\ 
&&\qquad\times\sum ^{l_\alpha}_{a=0}\,\sum ^{l_\beta}_{b=0}\ 
A_{_{\ell_\alpha N_\alpha;\ell_\beta N_\beta}}^{{\cal L},a,b}
\left( \frac{p_\alpha}{p_\beta}\right)^{a-b}\ ,
\end{eqnarray}
with $P_{\cal L}(x)$ the Legendre polynomial of order ${\cal L}$, and
\begin{equation}
x= \hat{p}_\alpha\cdot\hat{p}_\beta\qquad
\vec{q}_\alpha= -\vec{p}_\beta -\frac{1}{2} \vec{p}_\alpha
\qquad
\vec{q}_\beta =\vec{p}_\alpha +\frac{1}{2} \vec{p}_\beta\ .
\end{equation}
The coefficients $A_{_{\ell_\alpha N_\alpha;\ell_\beta 
N_\beta}}^{{\cal L},a,b}$ which results from the recoupling of the
spin and orbital angular momentum is given by
\begin{eqnarray}
\label{Acoef}
A_{_{\ell_\alpha N_\alpha;\ell_\beta N_\beta}}^{{\cal L},a,b} & = & (-1)^R \, 
\hat{\ell}_\alpha \, \hat{\ell}_\beta \, \hat{L}_\alpha \,
\hat{L}_\beta \, \hat{S}_\alpha \, \hat{S}_\beta \, 
\hat{\bar{\jmath}}_\alpha \, \hat{\bar{\jmath}}_\beta \, 
\hat{s}_\alpha \, \hat{s}_\beta \, \hat{\cal L}^2 \rho_{\alpha}^a 
\rho_{\beta}^b \nonumber \\
&&\times\sqrt{\frac{(2\ell_\alpha +1)!(2\ell_\beta +1)!}
{(2a)!(2b)!(2\ell_\alpha -2a)!(2\ell_\beta -2b)!}}\nonumber \\ 
&  & \times 
\sum _{f\Lambda \Lambda^\prime}
\left( \hat{f}\hat{\Lambda}\hat{\Lambda}^\prime \right) ^2 
\left\{ \begin{array}{ccc}
S_\alpha  & S_\beta  & f\\
L_\beta  & L_\alpha  & J 
\end{array}\right\}
\left\{ \begin{array}{ccc}
L_\alpha  & L_\beta  & f\\
\Lambda ^\prime  & \Lambda  & {\cal L}
\end{array}\right\}
\nonumber \\ && \times 
\left\{ \begin{array}{cccccccc}
j_\alpha & & S_\alpha  & & S_\beta & & j_\beta & \\
&\bar{\jmath}_\alpha & & f & & \bar{\jmath}_\beta & & j_\gamma \\
s_\alpha & & \ell_\alpha & & \ell_\beta & & s_\beta &
\end{array}\right\} \nonumber \\
&&\times
\left\{ \begin{array}{ccc}
\ell_\alpha  & \ell_\beta  & f\\
a & \ell_\beta -b & \Lambda \\
\ell_\alpha -a & b & \Lambda ^\prime  
\end{array}\right\} 
\left( \begin{array}{ccc}
a & \ell_\beta -b & \Lambda \\
0 & 0 & 0
\end{array}\right) \nonumber \\ &  & \times 
\left( \begin{array}{ccc}
\Lambda ^\prime  & {\cal L} & L_\beta \\
0 & 0 & 0
\end{array}\right) 
\left( \begin{array}{ccc}
\Lambda  & {\cal L} & L_\alpha \\
0 & 0 & 0
\end{array}\right) 
\left( \begin{array}{ccc}
\ell_\alpha -a & b & \Lambda ^\prime \\
0 & 0 & 0
\end{array}\right) \ ,\quad
\end{eqnarray}
where the $12-j$ symbol is that defined by Ord-Smith\cite{OS54},
the phase $R$ is defined as
\begin{eqnarray*}
R=-J+L_\alpha +L_\beta +S_\alpha +S_\beta +\bar{\jmath}_\alpha +
\bar{\jmath}_\beta -j_\alpha +s_\beta +\ell_\alpha +{\cal L}\ ,
\end{eqnarray*}
and finally $\rho_\alpha$ and $\rho_\beta$ are
\begin{eqnarray*}
\rho_\alpha =\frac{m_\beta}{m_\beta +m_\gamma}=\frac{1}{2}, \quad 
\rho_\beta =\frac{m_\alpha}{m_\alpha +m_\gamma}=\frac{1}{2}.
\end{eqnarray*} 

The isospin recoupling coefficient $B_{N_\alpha N_\beta}$ is given 
in terms of $6-j$ symbol by the relation
\begin{equation}
\label{Bcoef}
B_{_{N_\alpha N_\beta }}=(-1)^{\imath_\alpha +\imath_\gamma -
\bar{\imath}_\beta +2I}\, 
\hat{\bar{\imath}}_\alpha \, \hat{\bar{\imath}}_\beta \, 
\left\{ \begin{array}{ccc}
\imath_\beta & \imath_\gamma & \bar{\imath}_\alpha \\
\imath_\alpha & I & \bar{\imath}_\beta 
\end{array}\right\} .
\end{equation}

%Tables
%
\begin{table}[t]
\begin{center}
\caption{binding energy for a given potential and components of the 
wave function.}
\label{Table.1}
\begin{tabular}{ c  c  c  c  c  c } 
Potential & number of  & binding energy & $P(S)$ & $P(S')$ & $P(D)$ \\
& channels & (MeV)  & \% & \% & \% \\
\hline 
RSC & $5$ & $-7.15$ & 88.37\% & 1.88\% & 8.89\% \\
YAM4 & $5$ & $-9.12$ & 93.08\% & 1.58\% & 4.97\% \\
YAM7 & $5$ & $-8.05$ & 89.1\% & 1.59\% & 8.71\% \\
PEST & $5$ & $-7.27$ & 89.3\% & 1.88\% & 8.11\% \\
PEST & $10$ & $-7.10$ & 89.72\% & 1.71\% & 7.85\% \\
PEST & $18$ & $-7.32$& 89.56\% & 1.66\% & 8.07\% 
\end{tabular}
\end{center}
\end{table}
\begin{table}
\caption{Effective polarisation of the nucleons 
in $^3$He for various potentials.}
\label{Tpola1}
\begin{tabular}{ c | c c c c | c c c c }
 & \multicolumn{4}{c |}{$\sum P(X)$} & \multicolumn{4}{c }{$\int f(y)$} \\ 
 & $n^+$ & $n^-$ & $p^+$ & $p^-$ & $n^+$ & $n^-$ & $p^+$ & $p^-$\\
\hline
PEST & $93.97\%$ & $6.03\%$ & $48.96\%$ & $51.04\%$ & $93.62\%$ & 
$6.32\%$ & $48.98\%$ &$50.96\%$ \\ 
RSC & $93.45\%$ & $6.55\%$ & $48.83\%$ & $51.17\%$ & $92.92\%$ & 
$6.79\%$ & $48.76\%$ & $50.95\%$ \\ 
YAM7 & $93.66\%$ & $6.34\%$ & $48.81\%$ & $51.19\%$ & $93.25\%$ & 
$6.35\%$ & $48.69\%$ &$50.92\%$ 
\end{tabular}
\end{table}
\begin{table}
\caption{Effective polarisation of the nucleons in $^3$He and $^3$H,
with two-body interaction adjusted to produce the experimental 
binding energies.}
\label{Tpola2}
\begin{tabular}{ c | c c c c | c c c c }
 & \multicolumn{4}{c |}{$\sum P(X)$} & \multicolumn{4}{c }{$\int f(y)$} \\
 & $n^+$ & $n^-$ & $p^+$ & $p^-$ & $n^+$ & $n^-$ & $p^+$ & $p^-$ \\
\hline
$^3$He & $93.97\%$ & $6.03\%$ & $48.91\%$ & $51.09\%$ &  $93.73\%$ & 
$6.24\%$ & $48.94\%$ & $51.02\%$ \\ 
$^3$H & $93.45\%$ & $6.55\%$ & $48.85\%$ & $51.15\%$ & $93.86\%$ & 
$6.13\%$ & $48.89\%$ & $51.10\%$ 
\end{tabular}
\end{table}
%
%Figures
%
\newpage
\newpage
\newpage

\end{document}